\documentclass[authoryear,preprint,review,12pt]{elsarticle}

\usepackage{amssymb}
\usepackage{amsmath}
\usepackage[strings]{underscore}
\usepackage{hyperref}

\usepackage{lineno}

\begin{document}

%\linenumbers

\begin{frontmatter}



\title{{\bf The young Sun's XUV-activity as a constraint for lower CO$_2$-limits in the Earth's Archean atmosphere}}
%\title{Atmospheric losses reveal the histories of the young Sun and the Archean Earth's atmosphere}

%\title{High atmospheric carbon dioxide levels and low solar activity during the Earth's Archean}


\author{Colin P. Johnstone$^{*1}$, Helmut Lammer$^2$, Kristina G. Kislyakova$^{1,2}$, Manuel Scherf$^2$ \& Manuel G\"udel$^1$}

\address{
$^1$University of Vienna, Department of Astrophysics, T\"{u}rkenschanzstrasse 17, 1180 Vienna, Austria
$^2$Space Research Institute, Austrian Academy of Sciences, Graz, Austria
}
\begin{abstract}
Despite their importance for determining the evolution of the Earth's atmosphere and surface conditions, the evolutionary histories of the Earth's atmospheric CO$_2$  abundance during the Archean eon and the Sun's activity are poorly constrained. In this study, we {apply} a state-of-the-art physical model for the upper atmosphere of the {Archean} Earth to study the effects of different atmospheric CO$_2$/{N$_2$} mixing ratios and solar activity levels on the escape of the atmosphere to space. We find that unless CO$_2$ was a major constituent of the atmosphere during the Archean eon, enhanced heating of the thermosphere by the Sun's strong X-ray and ultraviolet radiation would have caused rapid escape to space. We derive lower limits on the atmospheric CO$_2$ abundance of approximately 40\% at 3.8~billion years ago, which is likely enough to counteract the faint young Sun and keep the Earth from being completely frozen. Furthermore, our results indicate that the Sun {was most likely} born as a slow {to} moderate {rotating young G-star} to prevent rapid escape, putting essential constraints on the Sun's activity evolution throughout the solar system's history. {In case that there were yet unknown cooling mechanisms present in the Archean atmosphere, this could reduce our CO$_2$ stability limits, and it would allow a more active Sun.}
\end{abstract}


%\item Earth Archean upper atmosphere studied using state-of-the-art models.
%\item High atmospheric CO$_2$ levels and low solar activity needed to prevent rapid escape to space.
%\item At least 40\% of atmospheric gas was CO$_2$ at 3.8 billion year ago.
%\item Greenhouse effect from CO$_2$ was strong enough to solve faint young Sun problem.
%\item The Sun must have been born as a slow rotator with low activity.


\end{frontmatter}

\textcopyright 2021. Licensed under the Creative Commons \href{https://creativecommons.org/licenses/by-nc-nd/4.0}{CC-BY-NC-ND}

\section{Introduction}

During the Archean, extending from approximately 3.8 to 2.5 billion years ago (Ga), the Sun's total luminosity was 75-82\% of its current value (\citealt{Gough81}), meaning that the Earth with its current atmosphere would have been too cold to possess liquid oceans (\citealt{SaganMullen72}). This is in contradiction to several lines of evidence that liquid water was already present {at the surface during the Hadean (4.56\,Ga to $\approx$\,4.0\,Ga\footnote{Please note that the Hadean eon is not strictly defined but 4.0\,Ga is often used as the boundary to the Archean eon. See, e.g., \citet{Zahnle2007}, and \citet{Goldblatt2010}.}) and Archean eons ($\approx$\,4.0\,Ga to 2.4\,Ga) \citep[e.g.,][]{Feulner12,catling2020archean} such as elevated $^{18}$O/$^{16}$O ratios in zircons from 4.3\,Ga \citep{Mojzsis2001,Wilde2001} and hydrothermal quartz with fluid inclusions of Archean sea water at 3.5\,Ga \citep{Foriel2004}}, and the problem is known as the Faint Young Sun Paradox. The most popular solutions involve a stronger greenhouse effect and many candidates have been proposed as the dominant greenhouse gas. {Methane produced by microbes could have been present at levels of $>10^{-4}$\,bar \citep[see, e.g.,][]{Robinson2019}.} Other suggestions involve higher N$_2$ surface pressures enhancing the greenhouse effect by broadening the absorption lines of greenhouse gases such as CO$_2$ (\citealt{Goldblatt09}) and the production of the strong greenhouse gas nitrous oxide by enhanced solar activity (\citealt{Airapetian16}). {However, N$_2$-dominated atmospheres with high surface pressures and negligible amounts of CO$_2$ as suggested by \citealt{Johnson18} are unlikely to build up due to high thermal escape and moreover one would expect a further increase of $\delta ^{15}$N, which is not observed in the present atmosphere (\citealt{Lammer18}, \citealt{Lammer19}; \citealt{Gebauer20}; \citealt{Sprosz21})}.

The most obvious {greenhouse} candidate is CO$_2$ and several studies have attempted to quantify how much CO$_2$ must have been present to solve the problem (\citealt{Kasting87}; \citealt{vonParis08}; \citealt{Kienert12}; \citealt{WolfToon13}; \citealt{Charnay13}; \citealt{Charnay17}; {\citealt{Charnay2020}}).
Geochemical measurements extend back to the late Archean and suggest that the CO$_2$ partial pressure (pCO$_2$) was likely significantly higher than in the modern atmosphere (\citealt{Rye95}; \citealt{Hessler04}; \citealt{Sheldon06}; \citealt{Driese11}; \citealt{Kanzaki15}) though significant uncertainty exists and no measurements for the early Archean or the Hadean are available (see below for a summary of these constraints). It has been suggested that due to the removal of mass by the solar wind, the early Sun was slightly more massive and therefore more luminous than it in now which could solve the faint young Sun problem (\citealt{Spalding18}).
However, upper limits on the mass loss rates of young Sun-like stars from radio observations and considerations of how rapidly the rotation rates of Sun-like stars decrease with age rule out this solution (\citealt{Gaidos00}; \citealt{Johnstone15a}; \citealt{Fichtinger17}).

During the Archean, the Sun emitted higher levels of X-ray {($< 10\,$nm)}, extreme ultraviolet {(EUV; 10\,--\,91.2\,nm))}, and far ultraviolet {(FUV; 91.2\,--\,200\,nm)} radiation (where we use `XUV' here to refer to the wavelength range 1-400~nm, {which includes the ultraviolet between 200\,--\,400\,nm}).
This radiation is important since it is absorbed high in the atmosphere where it drives heating and photochemistry {such as the absorption line of CO$_2$ at 89.9\,nm, and for N$_2$ at 79.6\,nm}  (e.g., \citealt{Roble87}; {\citealt{Rees89}; \citealt{Bauer2004}}).
The early Sun's XUV evolution is not known, with the major uncertainty being its early rotation rate (\citealt{Johnstone15a}; \citealt{Tu15}).
Rotation is important since it determines a star's XUV luminosity, with rapid rotators being more luminous than slow rotators (\citealt{Wright11}).
Observations of young stellar clusters show that the Sun could have been born with a rotation rate between a few times to a few tens of times its current value (\citealt{Bouvier14}), meaning that very different scenarios for the Sun's activity evolution in the first billion years are possible (\citealt{Tu15}).
If the Sun was born as a rapid rotator, it would have remained at its maximum activity level for several hundred million years; if the Sun was born as a slow rotator, its activity level would have decreased to moderate levels within the first $\sim$20 million years.

At higher levels of solar XUV radiation, the Earth's atmosphere is hotter and more expanded (\citealt{Tian08}; \citealt{Tian09}; \citealt{Kuramoto13}), leading to more rapid losses to space {(\citealt{Lammer08}; \citealt{Lichtenegger10}; \citealt{Lammer18})}.
Under the very high XUV flux that we would expect for the young Sun during the early Hadean, the Earth's modern atmosphere would be heated so much that it would escape to space at rates high enough to rapidly remove the entire atmosphere (\citealt{Johnstone19}).
Also important is the atmosphere's composition, which has changed significantly during the Earth's lifetime (\citealt{Lammer18}).
For example, high amounts of CO$_2$ in the atmosphere would lead to cooling of the upper atmosphere and reduced loss rates (\citealt{Kulikov07}; \citealt{Lichtenegger10}).
Although {the fractionation of the heavy Xe isotopes suggests some evidence for hydrogen escape from the Archean atmosphere (\citealt{Zahnle19})}, the very small atmospheric $^{14}$N/$^{15}$N fractionation compared to the deep mantle (\citealt{cartigny2013nitrogen}; \citealt{Fueri15}) {on the other hand} suggests only minor escape of nitrogen (\citealt{Lammer18}).

{N$_2$ likely has been outgassed and built up during the Archean eon (\citealt{Mikhail14}; \citealt{Som16}; \citealt{Lammer18}; \citealt{Lammer19}; \citealt{Stueken20}; \citealt{Sprosz21}), and became the main atmospheric species besides CO$_2$. The amount of N$_2$ present in the atmosphere when life evolved is important for understanding the production of prebiotic molecules (\citealt{Airapetian16}; \citealt{Zerkle17}; \citealt{ZerkleMikhail17};\citealt{Lammer18}; \citealt{Lammer19}; \citealt{Gebauer20}; \citealt{Sprosz21}).
However, estimates of atmospheric molecular CO$_2$ and N$_2$ partial surface pressures during the Archean eon widely vary in the literature.
So far previous studies focused on hydrodynamic hydrogen escape from early Earth' atmosphere during the Archean (\citealt{Tian05}; \citealt{Kuramoto13})
with the main aim to reproduce the observed fractionation of heavy Xe isotopes (\citealt{Zahnle19}; \citealt{Avice20}). While \citealt{Tian05} and
\citealt{Kuramoto13} neglected the atmospheric bulk gases, CO$_2$ and N$_2$, \citet{Zahnle19} developed a hydrodynamic diffusion-limited hydrogen escape model where he included CO$_2$ as the main atmospheric gas and considered CO$_2$-H$_2$-H atmospheres. These authors have chosen CO$_2$ rather than N$_2$ because CO$_2$ is
important for the energy budget in the thermosphere where its mixing has a great effect on the radiative cooling and heating (\citealt{Gordiets78}; \citealt{Gordiets82}; \citealt{Gordiets85}; \citealt{Kulikov06}; \citealt{Kulikov07}; \citealt{Johnstone18}; \citealt{Zahnle19}). Although \citet{Zahnle19} included
radiative cooling of CO$_2$ in their study of hydrogen escape from a CO$_2$-dominated atmosphere, they neglected dissociation of CO$_2$ molecules and escape of its dissociation products.}

{The aim of this particular study is to investigate various CO$_2$/N$_2$ mixing ratios and the escape of early Earth's bulk atmosphere during the Archean eon
exposed to a possible range of higher XUV fluxes from the young Sun as inferred from stellar rotation rates of solar-like stars (Tu et al. 2015). In Section 2, we summarize
the applied state-of-the-art physical-chemical thermosphere and escape model, and in Section 3 we present model results. In Section 4, we discuss the limitations
of the Archean CO$_2$-levels, and in Section 5 we address possible influences of minor atmospheric species that are not included in the present study.
Finally in Section 6, we address the implications of our findings for the faint young Sun problem and the young Sun's activity evolution before we
conclude our investigations in Section 7.}

%---------------------------------------------------------------------------------

\section{Upper atmosphere model}

We model the Earth's mesosphere and thermosphere using The Kompot Code, {which} is a sophisticated state-of-the-art physical {and chemical}
model for planetary {thermospheres} developed by \citet{Johnstone18} and \citet{Johnstone19} that considers
{heating from the absorption of solar X-ray, UV, and IR radiation, electron heating from collisions with non-thermal photoelectrons,
heating from exothermic reactions, Joule heating, radiative cooling from IR emission by several species,
thermal conduction, and energy exchange between the electron gas, ions and neutrals. For the
chemical structure, relevant chemical reactions, eddy and molecular diffusion, and advection are included.}
A major strength of this model is its comprehensive and first-principles treatment of the main physical processes including
a first principles treatment of the atmospheric heating that does not require the use of any free parameters
such as the commonly used heating efficiency. The model can successfully reproduce the upper atmospheric structures of
modern Venus, and Earth (\citealt{Johnstone18}). {The model also handles a possible transition from hydrostatic to hydrodynamic atmospheric conditions by
solving the hydrodynamic equations.}

In all simulations presented in this study, we calculate the 1D physical structure of the atmosphere between
the lower boundary at 50~km altitude and the upper boundary at the exobase {level where} the atmospheric gas becomes non-collisional.
The simulations start with a set of arbitrary initial conditions and evolve through a large number of time steps
until the atmospheres come to steady states. The exobase altitude is allowed to vary freely during the simulation.
Within the simulation domain, the gas is composed of 30 molecular and atomic species, including 11 ions.
{The reactions of the chemical network of the model can be seen in Table H.1. of \citet{Johnstone18}.
The neutral, ion, and electron components of the gas are assumed to have separate temperatures.
The composition of the gas {in the bulk atmosphere} at the lower boundary is assumed to be N$_2$ and CO$_2$,
with the relative mixing ratios of the two being {the most} important parameters that we study here.
{We will discuss the effects of possible existing minor species that are not included in this particular
study to our results in Section 5.}}

The lower boundary temperatures and particle densities are assumed to be 267~K and \mbox{$5 \times 10^{16}$~cm$^{-3}$} in all simulations.
These are approximately the values that we use in our modern Earth simulation (\citealt{Johnstone18}) and are to a large extent arbitrary.
Our results do not vary significantly with reasonable changes in the lower boundary temperature since what matters the most is the absorbed solar XUV energy, and our assumed temperature is approximately the planet's effective temperature which is a reasonable assumption.
The exact value of the base density is not important as long as it is sufficiently high that all of the input X-ray, EUV, and FUV radiation is absorbed in the simulation domain.
Making this value larger or smaller only has the effect of making small changes in the altitudes at which the various processes take place, which is unimportant since the base altitude in our simulation is anyway negligible compared to the radius of the Earth.

The thermal structure of the atmosphere is evolved in time by heating from several processes, cooling by infrared radiation to space by CO$_2$, NO, and O, thermal conduction applied to the neutral, ion, and electron components separately, and energy exchanges between these components by several elastic and inelastic collisional processes.
Our physical model for the heating of the gas is a particular strength since we do not assume any free parameters and calculate the heating rate from first principles.
Heating takes place due to the absorption of XUV radiation, which directly heats the gas, leads to the release of heat from exothermic chemical reactions {\citep[see also Fig.~5 and Table~H1 in][]{Johnstone18}}, and creates a spectrum of high energy electrons which heat the thermal electron gas through elastic collisions.
Additionally, the absorption of solar photospheric infrared radiation by CO$_2$ is included and can be important in simulations with large CO$_2$ mixing ratios.
Our model for CO$_2$ cooling makes no assumption about local thermodynamic equilibrium and is able to realistically calculate the cooling rates for very different CO$_2$ abundances, as can be seen from our calculations of modern Earth and Venus (\citealt{Johnstone18}).
Finally, in some cases in our model {simulations}, hydrodynamic advection driven by strong escape at the exobase is strong enough to cool the atmosphere significantly by adiabatic cooling.

{Since our model is a 1D model, we have to simplify radiation transfer. In \citet{Johnstone18}, we assumed that the computational domain is pointing in an arbitrary direction relative to the position of the star, with the angle between this direction and the direction that points directly to the star being defined as zenith angle $\theta$. We then calculated the XUV spectrum at each point in the atmosphere along this direction by doing the radiation transfer from the exobase to each point separately \citep[see Fig.~1 in][for a demonstration of this geometry]{Johnstone18} and found that for the Earth an angle of $\theta = 66^{\circ}$ gives the best representation of the atmosphere averaged over all zenith angles. We, therefore, also used $\theta = 66^{\circ}$ within our current study.}

\begin{figure}
\centering
\includegraphics[trim = 0mm 0mm 0mm 0mm, clip=true,width=0.7\textwidth]{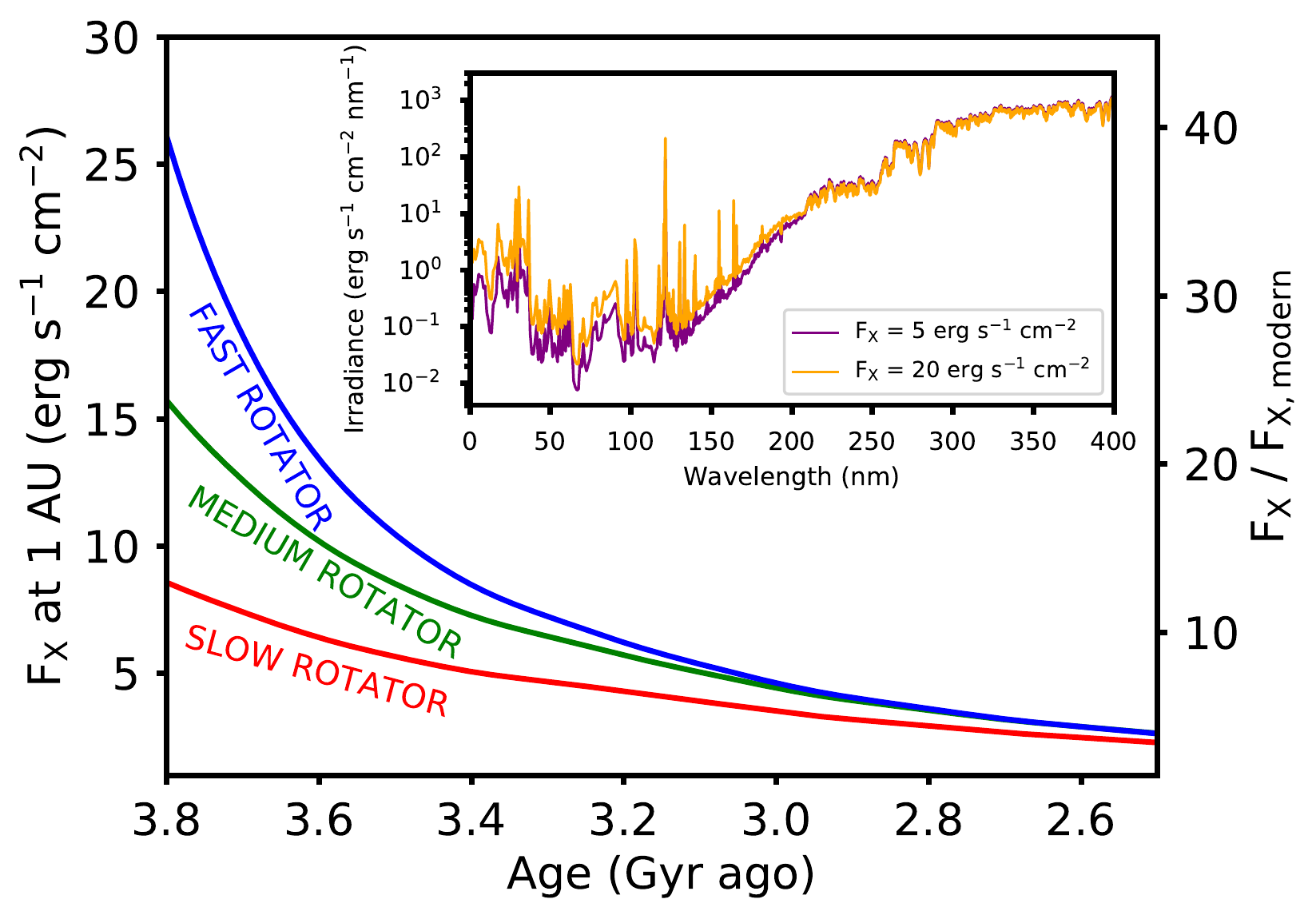}
\caption{
The evolution of the Sun's X-ray flux at the orbit of the Earth during the Archean for three cases of the Sun rotational evolution.
These cases are for the Sun being a slow (10th percentile), medium (50th percentile), and fast (90th percentile) rotator.
The initial rotation periods (at 1~Myr after solar system formation) for the Sun in these three scenarios are 9, 4, and 0.7 days respectively.
Two of the X-ray and ultraviolet spectra from \citet{Claire12} that we use at two different values for the X-ray flux are shown in the insert.
\label{fig:XUVevo}
}
\end{figure}

The chemical structure of the atmosphere is evolved in time by chemical reactions, molecular and eddy diffusion, and hydrodynamic advection. The chemical network used here is the network presented in \citet{Johnstone18}, with the only difference here being that we exclude all reactions that involve elements other than N, C, and O. The chemical network consists of 241 reactions in total, of which 42 are XUV driven photoionization or photodissociation reactions. Each of these photoreactions have their own wavelength dependent cross-sections. In Fig.~\ref{fig:XUVevo}, we show the evolution of the Sun's X-ray ($<$10~nm) flux at the orbit of the Earth during the Archean {eon assuming modern values for the orbital parameters}, where the different evolutionary tracks are for different cases for the Sun's rotational evolution (\citealt{Tu15}).
Our atmosphere model considers the full solar spectrum in the wavelength range \mbox{1-400~nm}, and for this spectrum, we use empirical models developed for use in atmospheric studies (\citealt{Claire12}), with two examples shown in Fig.~\ref{fig:XUVevo}. At each point on the X-ray evolution tracks, we assume the XUV spectrum from \citet{Claire12} with the same total X-ray flux.

{Here, we do not consider atmospheres that are heated to such high temperatures that they reach fully hydrodynamic states with the outflow velocities exceeding the escape velocity below the exobase. Although our model can be used for such cases} (\citealt{Johnstone19}; \citealt{Johnstone20}), {we find very high atmospheric loss rates of the assumed bulk atmospheric species can take place already before reaching the transonic hydrodynamic regime.}
As in \citet{Tian08} and \citet{Johnstone18}, we solve the hydrodynamic structure of the atmosphere assuming steady states for the total mass density and velocity structure (i.e. assuming \mbox{$\partial \rho / \partial t = 0$} and \mbox{$\partial v / \partial t = 0$}).
As described in \citet{Johnstone18}, we solve the full time-dependent hydrodynamic energy equations separately for the neutral, ion, and electron gases, with no steady state assumption {and the effects of adiabatic cooling of the gas due to the outward expansion are taken into account.}

{As seen below, we are interested here in which sets of atmospheric and solar parameters would lead to rapid expansion of the Earth's atmosphere driven by XUV heating, and for this purpose it is not necessary to consider each of the individual atmospheric loss mechanisms and we consider Jeans escape at the exobase only. \textbf{Considering hydrodynamic escape would not affect our results, since in such a case the atmosphere would anyway be lost completely.}}

\begin{figure}
\centering
\includegraphics[width=8cm]{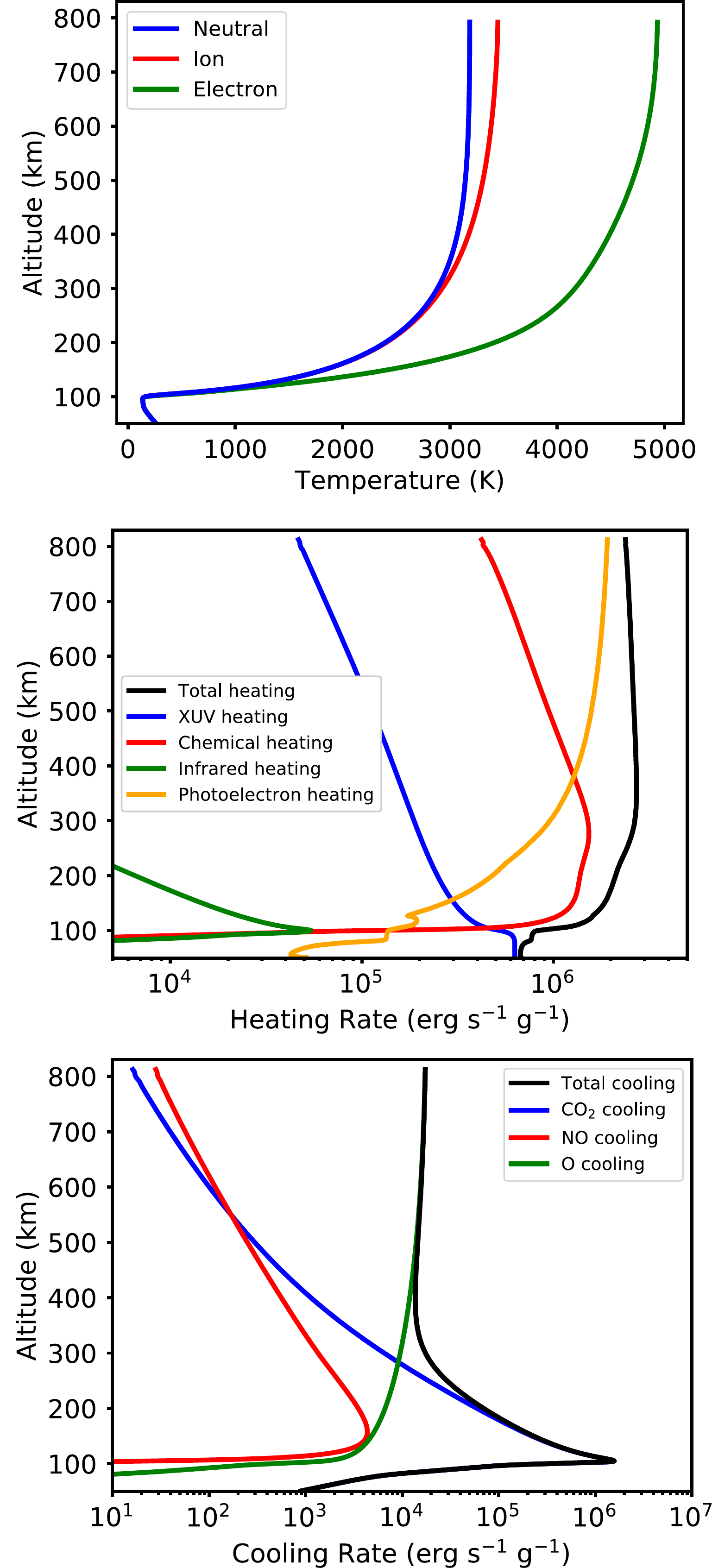}
\caption{
The vertical structures of temperature (\emph{upper-panel}), heating (\emph{middle-panel}), and cooling (\emph{lower-panel}) for our example simulation.
}
\label{fig:thermal}
\end{figure}

\begin{figure}
\centering
\includegraphics[trim = 0mm 0mm 0mm 0mm, clip=true,width=0.49\textwidth]{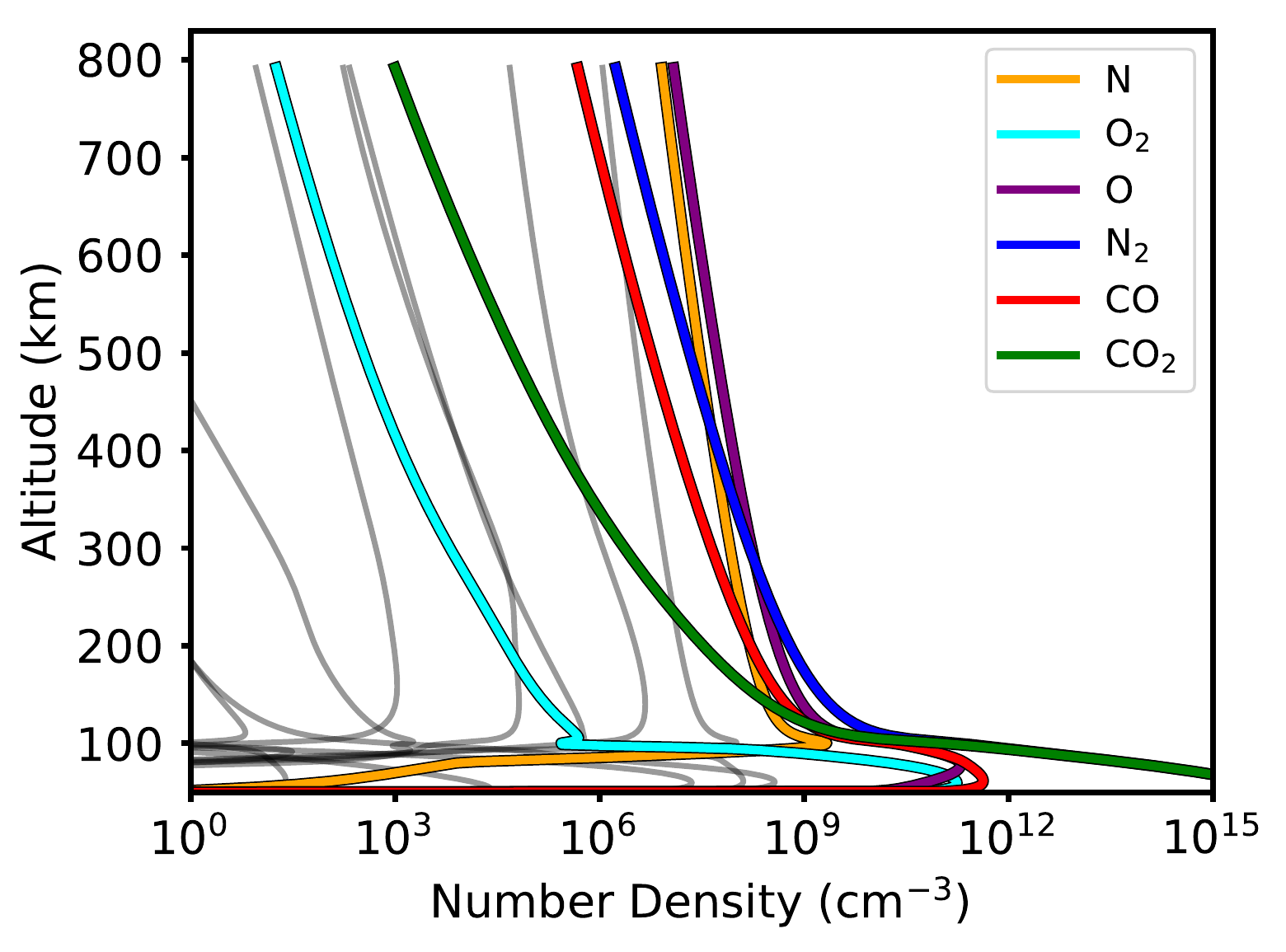}
\includegraphics[trim = 0mm 0mm 0mm 0mm, clip=true,width=0.49\textwidth]{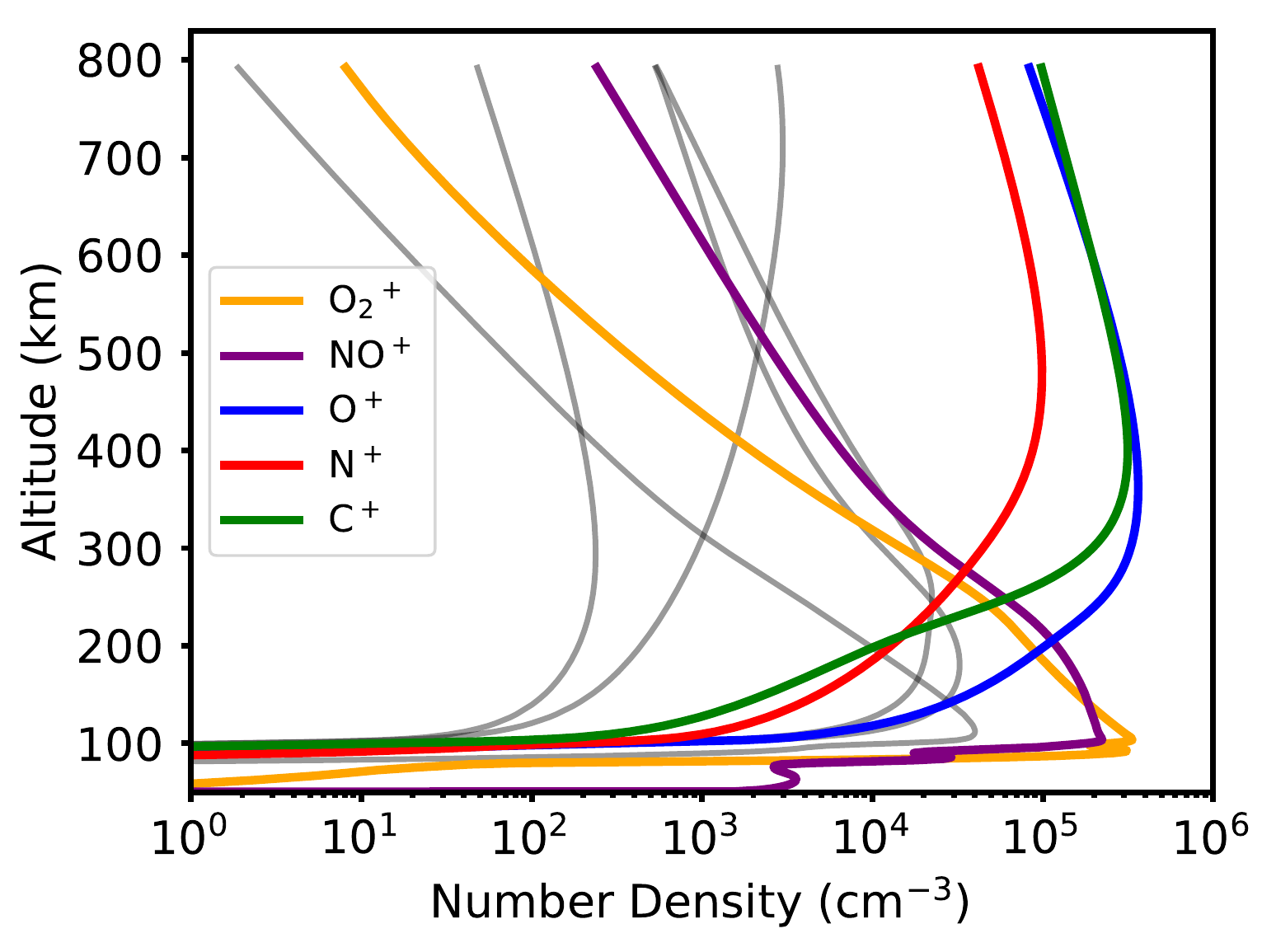}
\caption{
The vertical structures of neutral (\emph{left-panel}) and ion (\emph{right-panel}) species in our example simulation.
In both panels, several of the most important species are shown with coloured lines, as identified in the legend, and all the other species are shown as thin grey lines.
}
\label{fig:chemical}
\end{figure}

%---------------------------------------------------------------------------------

\section{CO$_2$/N$_2$ mixing ratios and the expansion of the thermosphere}

\begin{figure}
\centering
\includegraphics[trim = 0mm 0mm 0mm 0mm, clip=true,width=0.4\textwidth]{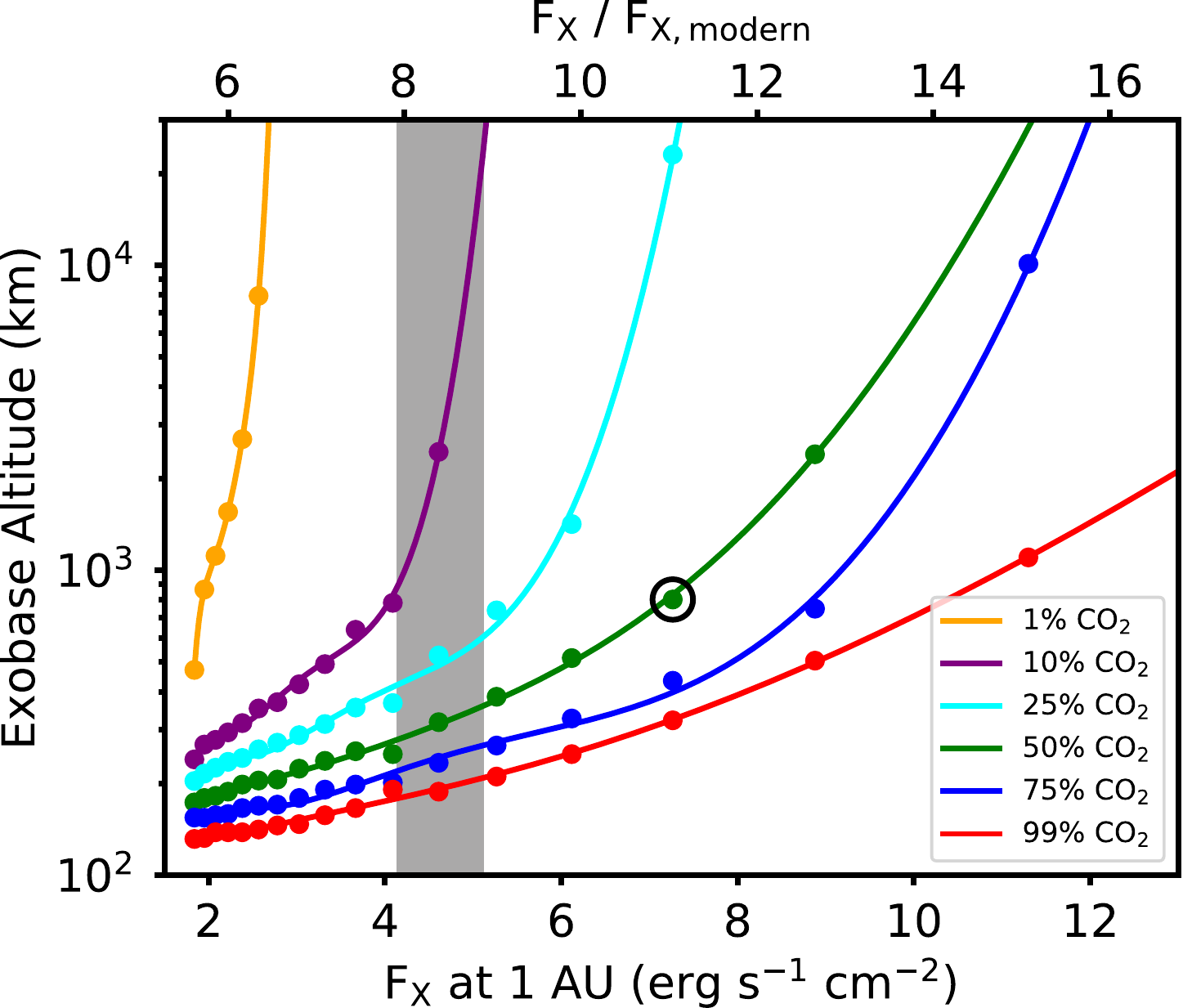}
\includegraphics[trim = 0mm 0mm 0mm 0mm, clip=true,width=0.4\textwidth]{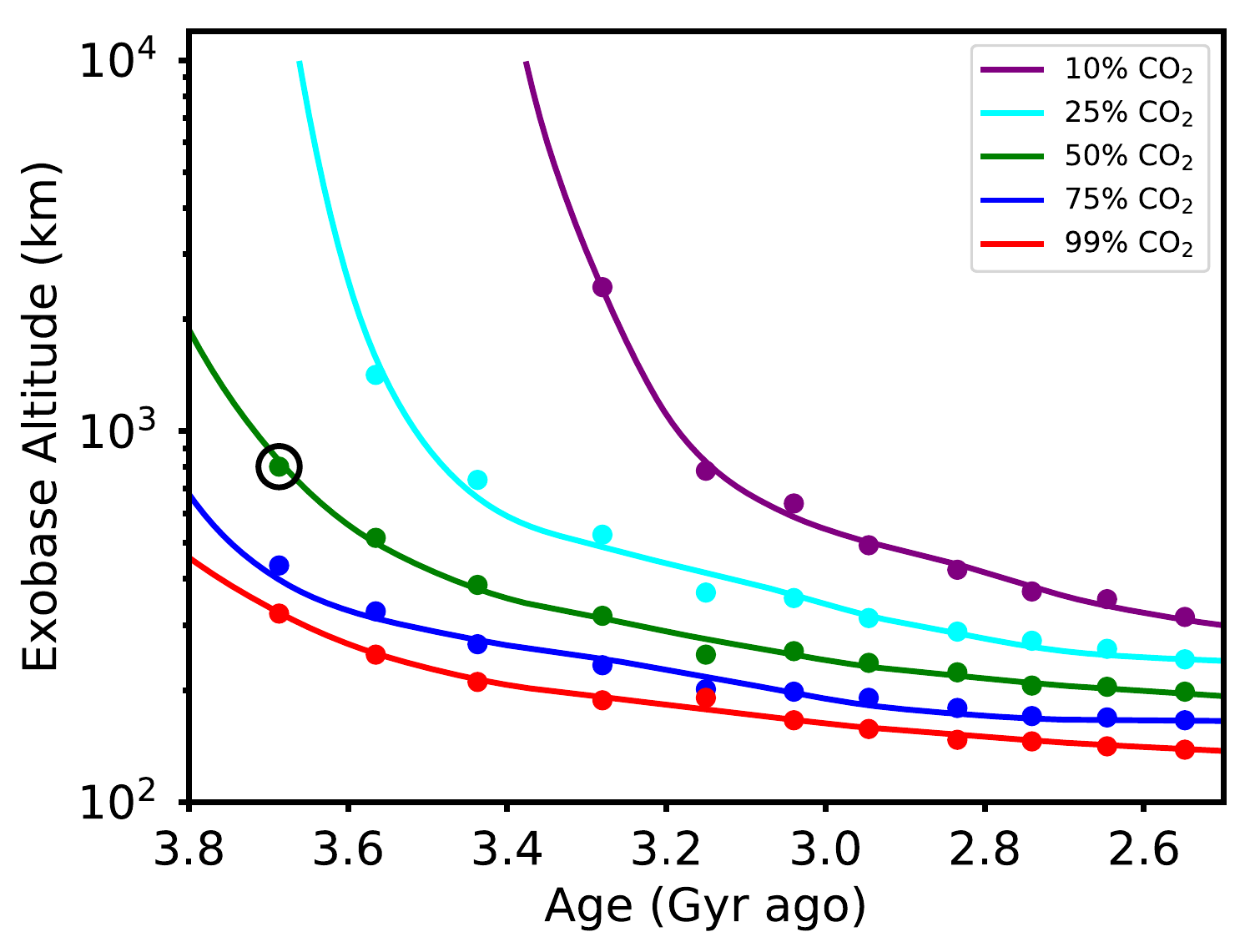}
\includegraphics[trim = 0mm 0mm 0mm 0mm, clip=true,width=0.4\textwidth]{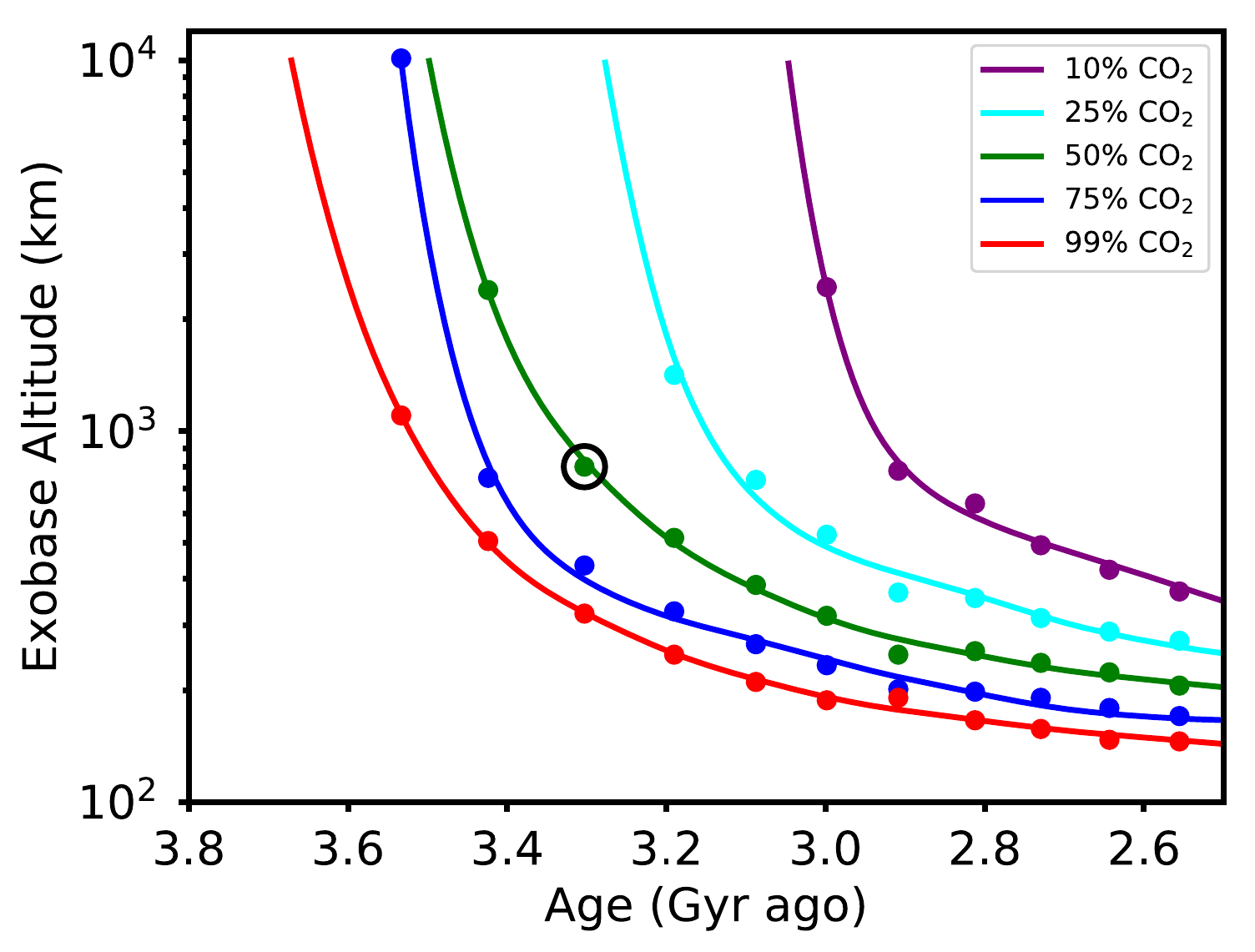}
\caption{
\emph{Upper panel:} the dependence of the exobase altitude on the X-ray flux at 1~au for different CO$_2$ mixing ratios.
\emph{Middle panel:} the evolution of the Earth's exobase altitude for cases of the Sun being a slow rotator for different CO$_2$ mixing ratios.
\emph{Lower panel:} the evolution of the Earth's exobase altitude for cases of the Sun being a fast rotator for different CO$_2$ mixing ratios.
In each case, circles show results of individual atmosphere models and the solid lines are derived from functional fits to the relationship between F$_\mathrm{X}$ and exobase altitude (shown in the upper panel) combined with the F$_\mathrm{X}$ evolution tracks. {The black circle depicts our example simulation.}
}
\label{fig:exoaltitude}
\end{figure}

Using our atmosphere model, we simulate the Earth's upper atmosphere throughout the Archean, from 3.8 to 2.5 Ga ago, assuming an atmosphere composed of CO$_2$ and N$_2$ {as the main species}. We first calculate a grid of upper atmosphere models with a range of solar activity levels and atmospheric CO$_2$ mixing ratios. At each activity level, we calculate models with CO$_2$ mixing ratios at the base of the simulation domain of 0.99, 0.75, 0.50, 0.25, 0.1, 0.01, and 0.001, and with N$_2$ making up the rest of the atmosphere. We consider solar XUV spectra having X-ray fluxes between approximately 2 to 15 erg~s$^{-1}$~cm$^{-2}$. It is important to note that for our simulations, the chemical composition of the atmosphere is more important than the total amount of atmosphere since the processes that we model take place at the densities where the XUV radiation is absorbed. Changes to the atmosphere's mass would lead to small changes in the altitudes of the lower boundaries of our simulation which are anyway negligible in comparison to the Earth's radius.
%--------------------------------------------------------------------------------------------------------------------------------------------
% The paragraph below was moved from the section before to this one, which fits better in the flow of the paper when one reads it
%--------------------------------------------------------------------------------------------------------------------------------------------
{As an example we show model results in Figs.~\ref{fig:thermal} and \ref{fig:chemical} of an atmosphere that is composed of 50\% CO$_2$ and 50\% N$_2$ for a solar XUV spectrum with an X-ray flux of 7.3\,erg\,s$^{-1}$~cm$^{-2}$ which is about 11 times the average present-day solar X-ray flux, i.e., 0.66\,erg\,s$^{-1}$~cm$^{-2}$ (with a minimum of 0.22\,erg\,s$^{-1}$~cm$^{-2}$, and a maximum of 2.82\,erg\,s$^{-1}$~cm$^{-2}$, see \citet{Judge03}).}
The vertical structure of the {atmospheric temperature} of this {example} is shown in {the upper panel of} Fig.~\ref{fig:thermal},
{heating and cooling rates are shown in the middle and lower panels, respectively. The corresponding density profiles for neutral (left) and ion species (right) are shown in Fig.~\ref{fig:chemical}}.
An interesting result that can be seen in Fig.~\ref{fig:thermal} is that the heating by the Sun's infrared spectrum is negligible compared to the XUV heating despite the very high CO$_2$ abundance. This is different to the cases of modern Venus and Mars in which infrared heating is very important (\citealt{Fox91}), and this difference is due simply to the fact that the XUV heating scales with the Sun's activity whereas the infrared heating (which depends on the Sun's photospheric emission) remains approximately constant in time.

The dependence of exobase altitude on X-ray flux at 1~AU and the evolution of the exobase altitude for the slow and fast solar rotator tracks for different CO$_2$ mixing ratios are shown in Fig.~\ref{fig:exoaltitude}. Higher X-ray fluxes lead to more heating and lower CO$_2$ mixing ratios lead to less cooling, meaning that the atmospheres in the high-activity and low-CO$_2$ cases are hotter and more expanded. \textbf{This also means that the transition between the collision-dominated to collisionless regime, i.e., the exobase level, occurs at higher altitudes, becomes less gravitationally bound and the upper atmosphere will be consequently lost to space more rapidly.}
{For high XUV fluxes the thermosphere can be heated to high temperatures and the upper atmosphere starts to expand very efficiently so that} losses at the exobase cause a large scale outward flow of material, and the resulting adiabatic cooling to some extent counters the strong heating (\citealt{Tian05}; \citealt{Tian08}; \citealt{MurrayClay09}).

Going backwards in time, for many of the CO$_2$ mixing ratio cases the expansion of the atmosphere becomes very rapid due to the stronger solar XUV emission.
In such cases, escape of the main atmospheric constituents at the exobase (atomic N, O, and C) by {thermal} escape becomes high enough to remove the entire atmosphere in a very short time period.
When specifically this takes place differs for each CO$_2$ mixing ratio and for different solar activity tracks.
Such rapid loss could not have taken place during the Archean since it would have meant no significant atmosphere could have been present meaning that only cases in which such rapid escape does not take place are likely to be realistic.
The most striking feature shown in Fig.~\ref{fig:exoaltitude} is that at ages before approximately 3.5~billion years ago, massive atmospheric expansion cannot be avoided in the early Archean {and the exobase level starts to expand upwards to above 10\,000\,km} if the Sun was a fast rotator even with very high concentrations of CO$_2$. {For the Hadean eon, however, not even a slow rotating Sun might have prevented the complete erosion of the atmosphere, if no yet unknown cooling agents had been present.}

%---------------------------------------------------------------------------------

\section{Limits on Archean CO$_2$-levels}

\begin{figure}
\centering
\includegraphics[trim = 0mm 0mm 0mm 0mm, clip=true,width=0.7\textwidth]{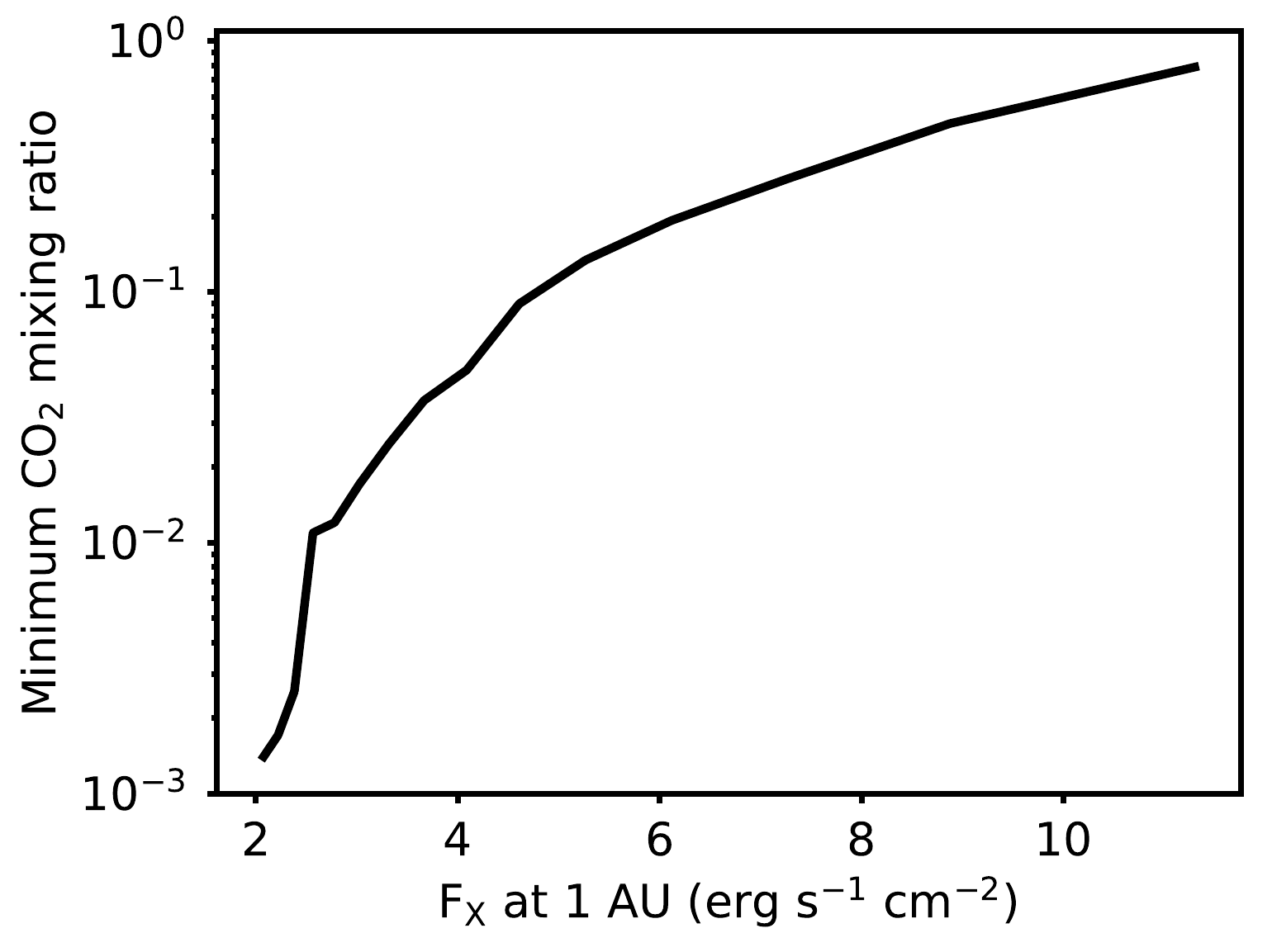}
\caption{
The minimum amount of CO$_2$ needed in the atmosphere to prevent rapid atmospheric escape as a function of the X-ray flux at 1 au.
Note that the entire solar X-ray and ultraviolet spectrum (1-400~nm) is considered in our models calculations.
}
\label{fig:CO2minLx}
\end{figure}

The fact that large amounts of atmospheric CO$_2$ are needed to prevent rapid escape allows us to derive restrictive lower limits on the Archean CO$_2$ content.
{This is based on the assumption that the Earth had a substantial atmosphere throughout the Archean, and there is significant geological evidence to support this assumption} (\citealt{catling2020archean}), {including evidence for liquid water and life being present during the early Archean and possibly before (\citealt{Mojzsis96}; \citealt{Javaux19})}.
To derive these lower limits, we first derive {them} as a function of activity level (see Fig.~\ref{fig:CO2minLx}), and then convert these limits into functions of age for the different solar activity tracks.
To do this, we define a threshold mass loss rate of 0.1~bar~Myr$^{-1}$ (the value at which a 1 bar atmosphere would be removed in {10} million years).
When the mass loss rate of the atmosphere is above this threshold, we consider this to be unrealistically large for the Archean.
Since our mass loss rates are very sensitive to the CO$_2$ mixing ratio, the threshold CO$_2$ mixing ratios are not sensitive to this exact definition for the threshold mass loss rate.
Assuming threshold mass loss rates of ten times larger or smaller would lead to almost identical results.
For each solar activity level, we use the model grid presented in the previous section and by extrapolation or interpolation we calculate the CO$_2$ mixing ratio at which the mass loss rate crosses {the} threshold value.
For many of the solar activity levels considered, we then tested the accuracy of our extrapolations/interpolations by running new models around the derived threshold CO$_2$ mixing ratio and found in each case that our original estimates were accurate.

Our lower limit on the CO$_2$ abundance naturally depends on the Sun's activity level, and therefore its initial rotation rate and age.
In Fig.~\ref{fig:CO2minLx}, we show the minimum CO$_2$ level needed to prevent rapid atmospheric escape as a function of the solar X-ray flux at 1~AU.
At a flux of 2~erg~s$^{-1}$~cm$^{-2}$, which is similar to that experienced by the Earth soon after the end of the Archean for all solar rotation scenarios, our lower limit is at a CO$_2$ mixing ratio of approximately $10^{-3}$.
At fluxes above approximately 13~erg~s$^{-1}$~cm$^{-2}$, rapid escape would take place even in a completely CO$_2$ atmosphere.
This is above the X-ray flux at the start of the Archean for the slow rotator case, and therefore CO$_2$ cooling is able to prevent rapid escape in this case, but below the fluxes from the medium and fast rotator cases.

In Fig.~\ref{fig:CO2evo}, we show how our lower limits on the atmospheric CO$_2$ mixing ratio evolves throughout the Archean for the three solar evolution cases, and compare these limits to geochemical measurements from the literature (\citealt{Rye95}; \citealt{Hessler04}; \citealt{Sheldon06}; \citealt{Driese11}; \citealt{Kanzaki15}).
Our results are consistent with all of the measurements if we assume a total surface pressure of $\sim$1~bar.
It has been argued that the mineralogy of Archaean sediments is inconsistent with a pCO$_2$ during the late Archean that is significantly higher than today (\citealt{Rosing10}), though this interpretation is a matter of debate (\citealt{Goldblatt11}; \citealt{Dauphas11}; \citealt{Reinhard11}).
Our results are inconsistent with such a low pCO$_2$ since rapid atmospheric escape would be unavoidable in this scenario.

%------------------------------------------------------------------------------------------
\section{Possible influences of minor atmospheric species}

{Besides the main atmospheric species CO$_2$ and N$_2$ there is much evidence that H$_2$O and biogenic CH$_4$ were most likely also present around 2.6\,--\,2.8\,Ga (\citealt{Hinrichs02}; \citealt{Zahnle10}; \citealt{Zerkle12}; \citealt{Lammer18}).
Below the cold trap for water vapor, H$_2$ will be produced from H$_2$O molecules by photolysis, and
photosynthesis followed by fermentation or diagenesis of organic matter. Highly reduced gases such as CH$_4$ or NH$_3$ are also dissociated
by photolysis due to the high XUV flux of the young Sun and by reactions with OH and O radicals produced from
the photolysis of H$_2$O and CO$_2$ molecules (\citealt{Kuhn79}; \citealt{Kasting82}; \citealt{Kasting93}; \citealt{Zahnle10}; \citealt{Wordsworth16}).
Above the cold trap, mainly the light dissociation products H$_2$, and H will be abundant, as discussed in detail by \citealt{Zahnle19}.}

{In agreement with our study, \citealt{Zahnle19} assumed that the bulk atmosphere during the Archean eon consisted mainly of CO$_2$ and N$_2$. However, these authors simplified their model by neglecting N$_2$ and considering only CO$_2$-H$_2$ atmospheres for the investigation of Xe escape.
\citealt{Zahnle19} developed a 1D hydrodynamic diffusion-limited hydrogen escape model that was applied to highly irradiated CO$_2$-H$_2$-H atmospheres.
Although their study includes CO$_2$-related radiative cooling, the CO$_2$-related photo- and ion-chemistry was also simplified, so that dissociation of CO$_2$ molecules was not included in their model. It was assumed that CO$_2$ does not escape but hydrogen diffuses through the CO$_2$ into the thermosphere and drags the embedded Xe$^+$ ions, which are considered as trace gas, upwards to the exobase from where it then escapes. It was shown in their study that the forces that act on the heavy Xe$^+$ ions are the collisions between the escaping neutral and ionized hydrogen that push Xe$^+$ ions outwards while collisions with CO$_2$ molecules block it. \citealt{Zahnle19} could reproduce Earth's present Xe isotope paradox if the assumed CO$_2$-dominated atmospheres had mixing ratios of at least 1\% hydrogen or 0.5\% CH$_4$ above the H$_2$O cold trap if they are exposed to a solar EUV irradiation that was $\geq$10 times higher than today's solar value.}

{Results very similar to those of \cite{Zahnle19} were also derived from hydro-code models that studied the losses caused by transonic escape of pure H$_2$ atmospheres; in those simulations, the lower boundary temperatures were fixed; they can also be seen as an in finite heat sink caused for instance by radiative cooling of CO$_2$. The differences that are attributable to the CO$_2$-related photochemistry that was included in the model of \citet{Zahnle19} but neglected by \citet{Tian05} and \citet{Kuramoto13} were more or less negligible. \citet{Zahnle19} also found that under their assumptions in the homosphere (lower thermosphere) where CO$_2$ is abundant, radiative heating and radiative cooling are in balance, the temperature of the gas is determined by the bulk gas and very little energy is channeled into hydrogen escape. This is also in agreement with a study of \citet{Yoshida20} who investigated the hydrodynamic hydrogen escape from an assumed reduced early Mars' atmosphere where they considered carbon compounds as radiative coolants.}

{From these results one can expect that the amount of hydrogen that is produced from the expected H$_2$O vapor and/or CH$_4$ molecules, which diffused through the water cold trap up to the exosphere was strong enough to drag heavy Xe$^+$ ions. However, the hydrogen flow does not contribute much to the adiabatic cooling that we obtain from our model simulations for expanded CO$_2$/N$_2$-dominated bulk atmospheres.
Therefore, one can expect that our heating and cooling rates shown in Fig.~\ref{fig:thermal}, as well as the resulting temperature and atmospheric profiles as shown in Figs.~\ref{fig:thermal} and \ref{fig:chemical} will not change much.}

{The hypothesis proposed by \cite{Zahnle19}, namely that hydrodynamic-diffusion-limited hydrogen escape is most likely the explanation for the observed Xe paradox, seems to be a solid one. However, when they neglect N$_2$ as a second main atmospheric species and neglect dissociation of CO$_2$ molecules and escape of its dissociation products, then their simplified model will not yield accurate results. As one can see from our study, different CO$_2$/N$_2$ mixing ratios exposed to various XUV fluxes
from the young Sun will yield differently expanded bulk atmosphere structures, and, hence, potentially high escape rates of C, O and N atoms.}

{For higher solar XUV fluxes as expected during the Hadean $> 4$\,Ga, other processes such as solar wind stripping and cold ion outflows of the bulk atmosphere could be important, which might be an additional challenge for sustaining a stable CO$_2$-dominated atmosphere. That ion outflow might have been significant during the Hadean eon is already indicated by recent findings of \citet{Kislyakova20} who studied the evolution of early Earth's polar outflow from the mid-Archean to present and already found loss rates due to polar outflow at $\approx$3 Ga of $\approx 3.3 \times 10^{27}$ s$^{-1}$ and $\approx 2.5 \times 10^{27}$ s$^{-1}$ for O$^+$ and N$^+$ ions. According to their results, the main parameters that governed the atmospheric escape during the studied time period are the evolution of the young Sun's XUV radiation and the atmosphere composition, while the evolution of the Earth's magnetic field had a less important role.}

{The cold trap, however, might not have existed continuously during the Archean eon due to fluctuations to very low total atmospheric surface pressures (\citealt{Som16}; \citealt{Lammer18}; \citealt{Stueken20}. If it did not exist, H$_2$O could have reached the thermosphere but would have subsequently been dissociated (\citealt{Zahnle16}). That such conditions at least periodically existed and H$_2$O was then indeed dissociated can be derived from $\approx$\,2.4\,Ga old micrometeorites which were oxidized in the upper atmosphere (\citealt{Tomkins16}; \citealt{Zahnle16}; \citealt{Rimmer20}). If H$_2$O was then indeed dissociated, it could not have contributed efficiently as cooling agent during this period.}

{Since our results have consequences for H$_2$ and the related Xe fractionation process,
future studies should redo the study by \citet{Zahnle19}
by including N$_2$ as a second major atmospheric constituent, CO$_2$ dissociation and the escape of C, O, and N atoms.
The observed Xe isotope ratio might then be reproduced more easily with lower XUV fluxes and even lower H$_2$ and/or CH$_4$ mixing ratios.}

%------------------------------------------------------------------------------------------

\section{The Faint Young Sun Paradox and the solar activity evolution}

\begin{figure}
\centering
\includegraphics[trim = 0mm 0mm 0mm 0mm, clip=true,width=0.7\textwidth]{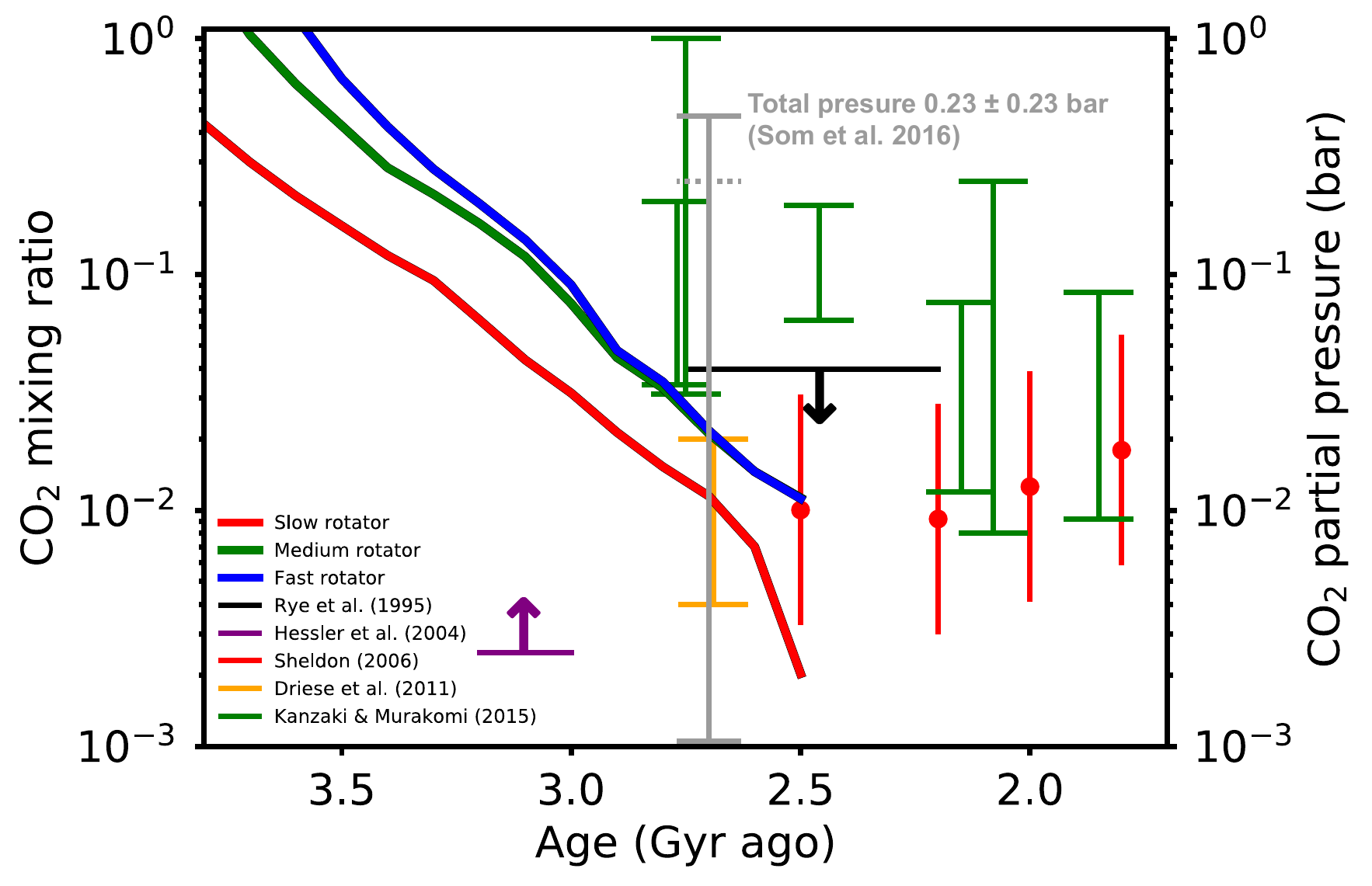}
\includegraphics[trim = 0mm 0mm 0mm 0mm, clip=true,width=0.7\textwidth]{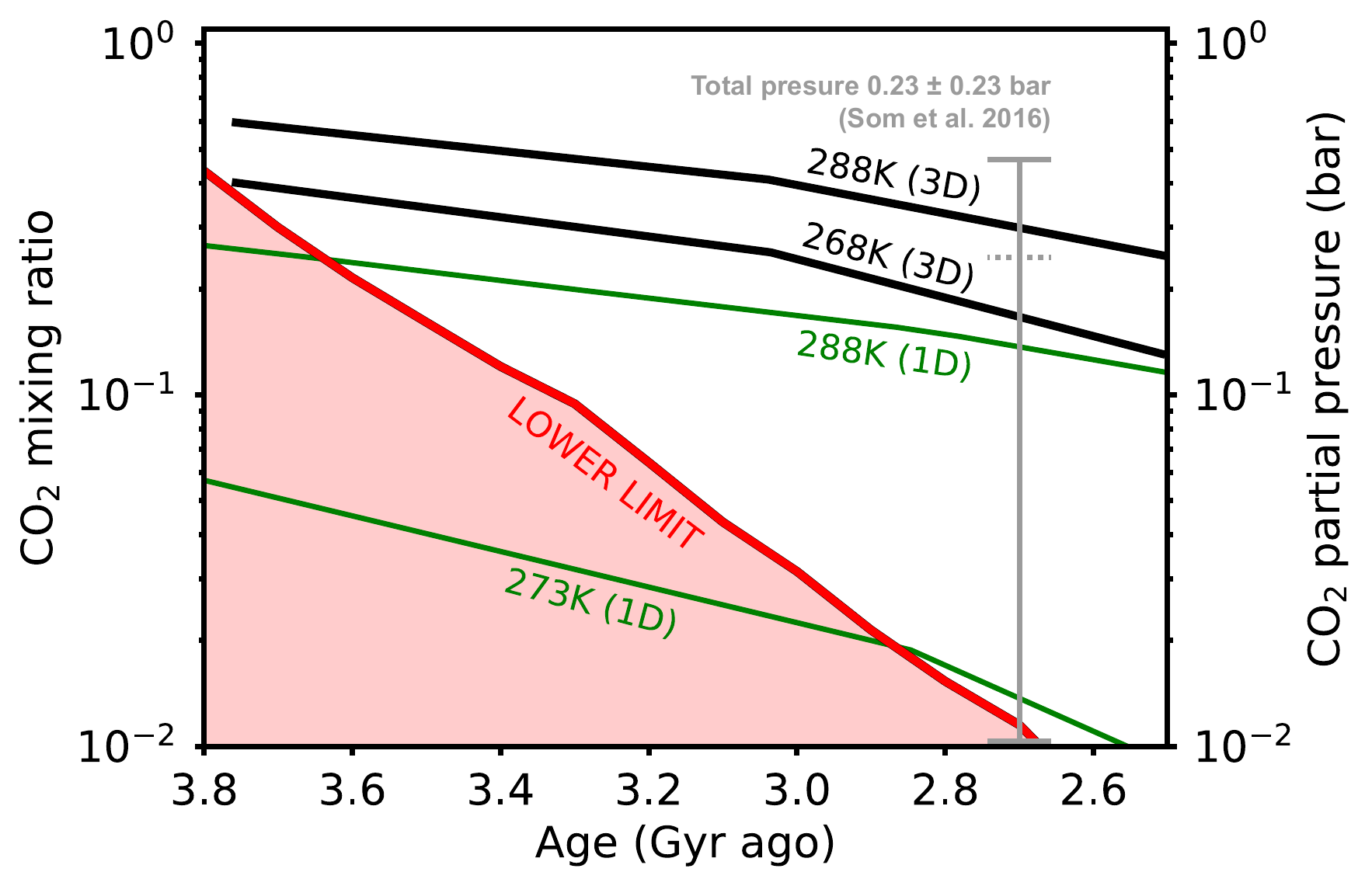}
\caption{
\emph{Upper-panel:} Our results for the minimum CO$_2$ mixing ratio throughout the Archean for the slow, medium, and fast rotating solar evolution scenarios.
The ranges and limits represent geochemical measurements from the literature.
\emph{Lower-panel:} A comparison of our lower limit on the CO$_2$ with estimates for the pCO$_2$ needed to keep the Archean Earth warm despite the fainter young Sun.
The red shaded area is ruled out by our results.
The green and black lines show respectively the estimates for pCO$_2$ required from 1D models (\citealt{vonParis08}) and 3D models (\citealt{Kienert12}), where the 3D models require a higher CO$_2$ concentration due to the ice-albedo feedback effect. {The grey lines indicate the total pressure at 2.7\,Ga of $0.23 \pm 0.32$\,bar as inferred from gas bubbles in basaltic lava flows by \citet{Som16}}.
}
\label{fig:CO2evo}
\end{figure}

Several studies have estimated the amount of CO$_2$ that would be needed during the Archean to solve the faint young Sun problem (\citealt{Kasting87}; \citealt{vonParis08}; \citealt{Kienert12}). Estimates using 3D global climate models that include ice-albedo feedback suggest that a very high amount of CO$_2$ of up to 0.4~bar are needed in the early Archean (\citealt{Kienert12}). Other 3D climate models suggest temperate climates can be achieved at 3.8~billion years ago with a pCO$_2$ of 0.1 to 0.36~bar {\citep{Charnay17,Charnay2020}}.
In Fig.~\ref{fig:CO2evo}, we compare our lower limit on the CO$_2$ mixing ratios for the slow solar rotator case to these results assuming again a total surface pressure of $\sim$1~bar. Our results show that the amount of CO$_2$ in the atmosphere during the early Archean was enough for the greenhouse effect from CO$_2$ alone to counter the faint young Sun.
%A surface pressure above 1~bar would strengthen our conclusions.

Later in the Archean, our results are less restrictive and we require, as a lower limit, less CO$_2$ to prevent rapid escape than is necessary to keep the surface warm.
It has been suggested that 2.5 billion years ago, a higher surface pressure of N$_2$ would have enhanced the greenhouse effect through pressure broadening of the absorption lines of greenhouse gases, and that with a pCO$_2$ of $10^{-2}$~bar at 2.5 billion years ago, a pN$_2$ of 2~bar would be needed to counter the fainter Sun (\citealt{Goldblatt09}).
In this scenario, the CO$_2$ mixing ratio is \mbox{$5 \times 10^{-3}$}, which is slightly above our lower limit at 2.5 billion years ago.
The situation is different at slightly earlier ages; this lower limit on the CO$_2$ mixing ratio is \mbox{$10^{-2}$} and \mbox{$2 \times 10^{-2}$} at 2.7 and 2.9 billion years ago respectively, suggesting that a higher pN$_2$ is unlikely to be realistic before the end of the Archean unless also accompanied by a correspondingly larger pCO$_2$.

We show in both Fig.~\ref{fig:exoaltitude} and Fig.~\ref{fig:CO2evo} that the Sun was most likely born as a slow {to intermediately rotating young G-star}.
No amount of CO$_2$ in the atmosphere would have prevented rapid escape to space had the Sun been more rapidly rotating.
This can be seen in Fig.~\ref{fig:CO2evo} where the minimum CO$_2$ mixing ratios in both the medium and fast rotator cases exceed unity.
By putting important constraints on the Sun's activity evolution, our results are also important for our understanding of the Earth's atmosphere evolution during the Hadean, and for the atmospheric evolutionary histories of Venus and Mars.
Since the Sun could not have been a fast rotator, its high level of activity must have decayed to low or moderate levels already very early in its lifetime, likely within the first 20-50 million years after the start of the solar system formation (\citealt{Tu15}) and was therefore not at its maximum activity level for the first 100~Myr as is commonly believed.
Such lower activity levels for slowly rotating stars can already be seen in the 12~million year old stellar cluster h~Per (\citealt{Argiroffi16}).

{The finding that the young Sun was most likely a slow to intermediately rotating young G-star also agrees with recent studies by Lammer et al. (2020a, 2020b, 2021) where Venus' and Earth's present atmospheric Ar, Ne isotope and bulk K/U ratios can be reproduced only if the young Sun was between a slow
and a moderately rotating young G-type star.}

It is interesting also to consider how important the slow rotation of the {young} Sun was for the formation of the habitable surface conditions of the Earth.
Had the Sun been a rapid rotator, our results suggest that the Earth's atmosphere would not have been able to survive during the Archean, and warm habitable surface conditions could not have been maintained.

%In this hypothetical scenario, it might still have been possible for the atmosphere to be replenished by outgassing from the mantle after the Sun's activity level declined and for habitable surface conditions similar to those of the modern Earth to form.
%Considering such scenarios might give us important information about the possibility of stars born as rapid rotators to possess Earth-like habitable planets.

%------------------------------------------------------------------------------------------
\section{Conclusions}

In this study, we use a state-of-the-art model for the Earth's upper atmosphere to put sensitive constraints on several parameters that are important for the evolution of the Earth's CO$_2$/N$_2$ atmosphere and surface conditions during the Archean.
These constraints come from considerations of the conditions necessary to prevent the {bulk} atmosphere from rapidly escaping to space during the Archean eon due to the {young} Sun's higher {X-ray and EUV} activity.
We derive lower limits on the amount of CO$_2$ that must have been present in the atmosphere to prevent such escape, finding that at the start of the Archean 3.8 billion years ago, {the atmospheric CO$_2$ level} must have been at least 40\% of the atmosphere.
This is important because no geochemical measurements of CO$_2$ from this time period or earlier are currently available and CO$_2$ is an important greenhouse gas.
Assuming an atmospheric surface pressure similar to that of the modern Earth, this is likely enough CO$_2$ to solve the faint young Sun problem during the early Archean.
Importantly, our results give constraints on the activity evolution of the young Sun which would have depended on its initial rotation rate.
Cooling by CO$_2$ is only able to prevent rapid atmospheric escape if the Sun was born as a slow rotator.
This result is very important not just for the Archean Earth, but for our understanding of atmospheric losses for the entire evolutionary history of the Earth and for
{terrestrial exoplanets as well.}

%------------------------------------------------------------------------------------------
\vspace{1cm}

\textbf{Acknowledgements}
C. Johnstone, H. Lammer, K. Kislyakova, and M. G\"udel, acknowledge support by the Austrian Science Fund (FWF) NFN project S11601-N16, ``Pathways to Habitability: From Disks to Active Stars, Planets and Life'' and the related FWF NFN subprojects, S11604-N16 ``Radiation \& Wind Evolution from the T Tauri Phase to ZAMS \& Beyond'', and S11607-N16 ``Particle/Radiative Interactions with Upper Atmospheres of Planetary Bodies under Extreme Stellar Conditions''. {We acknowledge one anonymous referee and Jeff Linsky for their comments which helped to improve the article.}



\bibliographystyle{elsarticle-harv}
\bibliography{mybib}

\begin{thebibliography}{85}
\expandafter\ifx\csname natexlab\endcsname\relax\def\natexlab#1{#1}\fi
\providecommand{\url}[1]{\texttt{#1}}
\providecommand{\href}[2]{#2}
\providecommand{\path}[1]{#1}
\providecommand{\DOIprefix}{doi:}
\providecommand{\ArXivprefix}{arXiv:}
\providecommand{\URLprefix}{URL: }
\providecommand{\Pubmedprefix}{pmid:}
\providecommand{\doi}[1]{\href{http://dx.doi.org/#1}{\path{#1}}}
\providecommand{\Pubmed}[1]{\href{pmid:#1}{\path{#1}}}
\providecommand{\bibinfo}[2]{#2}
\ifx\xfnm\relax \def\xfnm[#1]{\unskip,\space#1}\fi
%Type = Article
\bibitem[{{Airapetian} et~al.(2016){Airapetian}, {Glocer}, {Gronoff},
  {H{\'e}brard} and {Danchi}}]{Airapetian16}
\bibinfo{author}{{Airapetian}, V.S.}, \bibinfo{author}{{Glocer}, A.},
  \bibinfo{author}{{Gronoff}, G.}, \bibinfo{author}{{H{\'e}brard}, E.},
  \bibinfo{author}{{Danchi}, W.}, \bibinfo{year}{2016}.
\newblock \bibinfo{title}{{Prebiotic chemistry and atmospheric warming of early
  Earth by an active young Sun}}.
\newblock \bibinfo{journal}{Nature Geoscience} \bibinfo{volume}{9},
  \bibinfo{pages}{452--455}.
\newblock \DOIprefix\doi{10.1038/ngeo2719}.
%Type = Article
\bibitem[{{Argiroffi} et~al.(2016){Argiroffi}, {Caramazza}, {Micela},
  {Sciortino}, {Moraux}, {Bouvier} and {Flaccomio}}]{Argiroffi16}
\bibinfo{author}{{Argiroffi}, C.}, \bibinfo{author}{{Caramazza}, M.},
  \bibinfo{author}{{Micela}, G.}, \bibinfo{author}{{Sciortino}, S.},
  \bibinfo{author}{{Moraux}, E.}, \bibinfo{author}{{Bouvier}, J.},
  \bibinfo{author}{{Flaccomio}, E.}, \bibinfo{year}{2016}.
\newblock \bibinfo{title}{{Supersaturation and activity-rotation relation in
  PMS stars: the young cluster h Persei}}.
\newblock \bibinfo{journal}{\aap} \bibinfo{volume}{589}, \bibinfo{pages}{A113}.
\newblock \DOIprefix\doi{10.1051/0004-6361/201526539},
  \href{http://arxiv.org/abs/1602.03696}{{\tt arXiv:1602.03696}}.
%Type = Article
\bibitem[{{Avice} and {Marty}(2020)}]{Avice20}
\bibinfo{author}{{Avice}, G.}, \bibinfo{author}{{Marty}, B.},
  \bibinfo{year}{2020}.
\newblock \bibinfo{title}{{Perspectives on Atmospheric Evolution from Noble Gas
  and Nitrogen Isotopes on Earth, Mars \& Venus}}.
\newblock \bibinfo{journal}{\ssr} \bibinfo{volume}{216}, \bibinfo{pages}{36}.
\newblock \DOIprefix\doi{10.1007/s11214-020-00655-0},
  \href{http://arxiv.org/abs/2003.11431}{{\tt arXiv:2003.11431}}.
%Type = Book
\bibitem[{{Bauer} and {Lammer}(2004)}]{Bauer2004}
\bibinfo{author}{{Bauer}, S.J.}, \bibinfo{author}{{Lammer}, H.},
  \bibinfo{year}{2004}.
\newblock \bibinfo{title}{{Planetary aeronomy : atmosphere environments in
  planetary systems}}.
%Type = Inproceedings
\bibitem[{{Bouvier} et~al.(2014){Bouvier}, {Matt}, {Mohanty}, {Scholz},
  {Stassun} and {Zanni}}]{Bouvier14}
\bibinfo{author}{{Bouvier}, J.}, \bibinfo{author}{{Matt}, S.P.},
  \bibinfo{author}{{Mohanty}, S.}, \bibinfo{author}{{Scholz}, A.},
  \bibinfo{author}{{Stassun}, K.G.}, \bibinfo{author}{{Zanni}, C.},
  \bibinfo{year}{2014}.
\newblock \bibinfo{title}{{Angular Momentum Evolution of Young Low-Mass Stars
  and Brown Dwarfs: Observations and Theory}}, in: \bibinfo{editor}{{Beuther},
  H.}, \bibinfo{editor}{{Klessen}, R.S.}, \bibinfo{editor}{{Dullemond}, C.P.},
  \bibinfo{editor}{{Henning}, T.} (Eds.), \bibinfo{booktitle}{Protostars and
  Planets VI}, p. \bibinfo{pages}{433}.
\newblock \DOIprefix\doi{10.2458/azu_uapress_9780816531240-ch019},
  \href{http://arxiv.org/abs/1309.7851}{{\tt arXiv:1309.7851}}.
%Type = Article
\bibitem[{Cartigny and Marty(2013)}]{cartigny2013nitrogen}
\bibinfo{author}{Cartigny, P.}, \bibinfo{author}{Marty, B.},
  \bibinfo{year}{2013}.
\newblock \bibinfo{title}{Nitrogen isotopes and mantle geodynamics: The
  emergence of life and the atmosphere--crust--mantle connection}.
\newblock \bibinfo{journal}{Elements} \bibinfo{volume}{9},
  \bibinfo{pages}{359--366}.
%Type = Article
\bibitem[{Catling and Zahnle(2020)}]{catling2020archean}
\bibinfo{author}{Catling, D.C.}, \bibinfo{author}{Zahnle, K.J.},
  \bibinfo{year}{2020}.
\newblock \bibinfo{title}{The archean atmosphere}.
\newblock \bibinfo{journal}{Science Advances} \bibinfo{volume}{6},
  \bibinfo{pages}{eaax1420}.
%Type = Article
\bibitem[{{Charnay} et~al.(2013){Charnay}, {Forget}, {Wordsworth}, {Leconte},
  {Millour}, {Codron} and {Spiga}}]{Charnay13}
\bibinfo{author}{{Charnay}, B.}, \bibinfo{author}{{Forget}, F.},
  \bibinfo{author}{{Wordsworth}, R.}, \bibinfo{author}{{Leconte}, J.},
  \bibinfo{author}{{Millour}, E.}, \bibinfo{author}{{Codron}, F.},
  \bibinfo{author}{{Spiga}, A.}, \bibinfo{year}{2013}.
\newblock \bibinfo{title}{{Exploring the faint young Sun problem and the
  possible climates of the Archean Earth with a 3-D GCM}}.
\newblock \bibinfo{journal}{Journal of Geophysical Research (Atmospheres)}
  \bibinfo{volume}{118}, \bibinfo{pages}{10,414--10,431}.
\newblock \DOIprefix\doi{10.1002/jgrd.50808},
  \href{http://arxiv.org/abs/1310.4286}{{\tt arXiv:1310.4286}}.
%Type = Article
\bibitem[{{Charnay} et~al.(2017){Charnay}, {Le Hir}, {Fluteau}, {Forget} and
  {Catling}}]{Charnay17}
\bibinfo{author}{{Charnay}, B.}, \bibinfo{author}{{Le Hir}, G.},
  \bibinfo{author}{{Fluteau}, F.}, \bibinfo{author}{{Forget}, F.},
  \bibinfo{author}{{Catling}, D.C.}, \bibinfo{year}{2017}.
\newblock \bibinfo{title}{{A warm or a cold early Earth? New insights from a
  3-D climate-carbon model}}.
\newblock \bibinfo{journal}{Earth and Planetary Science Letters}
  \bibinfo{volume}{474}, \bibinfo{pages}{97--109}.
\newblock \DOIprefix\doi{10.1016/j.epsl.2017.06.029},
  \href{http://arxiv.org/abs/1706.06842}{{\tt arXiv:1706.06842}}.
%Type = Article
\bibitem[{{Charnay} et~al.(2020){Charnay}, {Wolf}, {Marty} and
  {Forget}}]{Charnay2020}
\bibinfo{author}{{Charnay}, B.}, \bibinfo{author}{{Wolf}, E.T.},
  \bibinfo{author}{{Marty}, B.}, \bibinfo{author}{{Forget}, F.},
  \bibinfo{year}{2020}.
\newblock \bibinfo{title}{{Is the Faint Young Sun Problem for Earth Solved?}}
\newblock \bibinfo{journal}{\ssr} \bibinfo{volume}{216}, \bibinfo{pages}{90}.
\newblock \DOIprefix\doi{10.1007/s11214-020-00711-9},
  \href{http://arxiv.org/abs/2006.06265}{{\tt arXiv:2006.06265}}.
%Type = Article
\bibitem[{{Claire} et~al.(2012){Claire}, {Sheets}, {Cohen}, {Ribas}, {Meadows}
  and {Catling}}]{Claire12}
\bibinfo{author}{{Claire}, M.W.}, \bibinfo{author}{{Sheets}, J.},
  \bibinfo{author}{{Cohen}, M.}, \bibinfo{author}{{Ribas}, I.},
  \bibinfo{author}{{Meadows}, V.S.}, \bibinfo{author}{{Catling}, D.C.},
  \bibinfo{year}{2012}.
\newblock \bibinfo{title}{{The Evolution of Solar Flux from 0.1 nm to 160
  {\ensuremath{\mu}}m: Quantitative Estimates for Planetary Studies}}.
\newblock \bibinfo{journal}{\apj} \bibinfo{volume}{757}, \bibinfo{pages}{95}.
\newblock \DOIprefix\doi{10.1088/0004-637X/757/1/95}.
%Type = Article
\bibitem[{{Dauphas} and {Kasting}(2011)}]{Dauphas11}
\bibinfo{author}{{Dauphas}, N.}, \bibinfo{author}{{Kasting}, J.F.},
  \bibinfo{year}{2011}.
\newblock \bibinfo{title}{{Low p$_{CO<SUB>2}$</SUB> in the pore water, not in
  the Archean air}}.
\newblock \bibinfo{journal}{\nat} \bibinfo{volume}{474}, \bibinfo{pages}{E1}.
\newblock \DOIprefix\doi{10.1038/nature09960}.
%Type = Article
\bibitem[{{Driese} et~al.(2011){Driese}, {Jirsa}, {Ren}, {Brantley}, {Sheldon},
  {Parker} and {Schmitz}}]{Driese11}
\bibinfo{author}{{Driese}, S.G.}, \bibinfo{author}{{Jirsa}, M.A.},
  \bibinfo{author}{{Ren}, M.}, \bibinfo{author}{{Brantley}, S.L.},
  \bibinfo{author}{{Sheldon}, N.D.}, \bibinfo{author}{{Parker}, D.},
  \bibinfo{author}{{Schmitz}, M.}, \bibinfo{year}{2011}.
\newblock \bibinfo{title}{{Neoarchean paleoweathering of tonalite and
  metabasalt: Implications for reconstructions of 2.69Ga early terrestrial
  ecosystems and paleoatmospheric chemistry}}.
\newblock \bibinfo{journal}{Precambrian Research} \bibinfo{volume}{189},
  \bibinfo{pages}{1--17}.
\newblock \DOIprefix\doi{10.1016/j.precamres.2011.04.003}.
%Type = Article
\bibitem[{{Feulner}(2012)}]{Feulner12}
\bibinfo{author}{{Feulner}, G.}, \bibinfo{year}{2012}.
\newblock \bibinfo{title}{{The faint young Sun problem}}.
\newblock \bibinfo{journal}{Reviews of Geophysics} \bibinfo{volume}{50},
  \bibinfo{pages}{RG2006}.
\newblock \DOIprefix\doi{10.1029/2011RG000375},
  \href{http://arxiv.org/abs/1204.4449}{{\tt arXiv:1204.4449}}.
%Type = Article
\bibitem[{{Fichtinger} et~al.(2017){Fichtinger}, {G{\"u}del}, {Mutel},
  {Hallinan}, {Gaidos}, {Skinner}, {Lynch} and {Gayley}}]{Fichtinger17}
\bibinfo{author}{{Fichtinger}, B.}, \bibinfo{author}{{G{\"u}del}, M.},
  \bibinfo{author}{{Mutel}, R.L.}, \bibinfo{author}{{Hallinan}, G.},
  \bibinfo{author}{{Gaidos}, E.}, \bibinfo{author}{{Skinner}, S.L.},
  \bibinfo{author}{{Lynch}, C.}, \bibinfo{author}{{Gayley}, K.G.},
  \bibinfo{year}{2017}.
\newblock \bibinfo{title}{{Radio emission and mass loss rate limits of four
  young solar-type stars}}.
\newblock \bibinfo{journal}{\aap} \bibinfo{volume}{599}, \bibinfo{pages}{A127}.
\newblock \DOIprefix\doi{10.1051/0004-6361/201629886},
  \href{http://arxiv.org/abs/1702.08393}{{\tt arXiv:1702.08393}}.
%Type = Article
\bibitem[{{Foriel} et~al.(2004){Foriel}, {Philippot}, {Rey}, {Somogyi}, {Banks}
  and {M{\'e}nez}}]{Foriel2004}
\bibinfo{author}{{Foriel}, J.}, \bibinfo{author}{{Philippot}, P.},
  \bibinfo{author}{{Rey}, P.}, \bibinfo{author}{{Somogyi}, A.},
  \bibinfo{author}{{Banks}, D.}, \bibinfo{author}{{M{\'e}nez}, B.},
  \bibinfo{year}{2004}.
\newblock \bibinfo{title}{{Biological control of Cl/Br and low sulfate
  concentration in a 3.5-Gyr-old seawater from North Pole, Western Australia}}.
\newblock \bibinfo{journal}{Earth and Planetary Science Letters}
  \bibinfo{volume}{228}, \bibinfo{pages}{451--463}.
\newblock \DOIprefix\doi{10.1016/j.epsl.2004.09.034}.
%Type = Article
\bibitem[{{Fox} and {Bougher}(1991)}]{Fox91}
\bibinfo{author}{{Fox}, J.L.}, \bibinfo{author}{{Bougher}, S.W.},
  \bibinfo{year}{1991}.
\newblock \bibinfo{title}{{Structure, luminosity, and dynamics of the Venus
  thermosphere}}.
\newblock \bibinfo{journal}{\ssr} \bibinfo{volume}{55},
  \bibinfo{pages}{357--489}.
\newblock \DOIprefix\doi{10.1007/BF00177141}.
%Type = Article
\bibitem[{{F{\"u}ri} and {Marty}(2015)}]{Fueri15}
\bibinfo{author}{{F{\"u}ri}, E.}, \bibinfo{author}{{Marty}, B.},
  \bibinfo{year}{2015}.
\newblock \bibinfo{title}{{Nitrogen isotope variations in the Solar System}}.
\newblock \bibinfo{journal}{Nature Geoscience} \bibinfo{volume}{8},
  \bibinfo{pages}{515--522}.
\newblock \DOIprefix\doi{10.1038/ngeo2451}.
%Type = Article
\bibitem[{{Gaidos} et~al.(2000){Gaidos}, {G{\"u}del} and {Blake}}]{Gaidos00}
\bibinfo{author}{{Gaidos}, E.J.}, \bibinfo{author}{{G{\"u}del}, M.},
  \bibinfo{author}{{Blake}, G.A.}, \bibinfo{year}{2000}.
\newblock \bibinfo{title}{{The Faint Young Sun Paradox: An observational test
  of an alternative solar model}}.
\newblock \bibinfo{journal}{\grl} \bibinfo{volume}{27},
  \bibinfo{pages}{501--503}.
\newblock \DOIprefix\doi{10.1029/1999GL010740}.
%Type = Article
\bibitem[{{Gebauer} et~al.(2020){Gebauer}, {Grenfell}, {Lammer}, {de Vera},
  {Spro{\ss}}, {Airapetian}, {Sinnhuber} and {Rauer}}]{Gebauer20}
\bibinfo{author}{{Gebauer}, S.}, \bibinfo{author}{{Grenfell}, J.L.},
  \bibinfo{author}{{Lammer}, H.}, \bibinfo{author}{{de Vera}, J.P.P.},
  \bibinfo{author}{{Spro{\ss}}, L.}, \bibinfo{author}{{Airapetian}, V.S.},
  \bibinfo{author}{{Sinnhuber}, M.}, \bibinfo{author}{{Rauer}, H.},
  \bibinfo{year}{2020}.
\newblock \bibinfo{title}{{Atmospheric Nitrogen When Life Evolved on Earth}}.
\newblock \bibinfo{journal}{Astrobiology} \bibinfo{volume}{20},
  \bibinfo{pages}{1413--1426}.
\newblock \DOIprefix\doi{10.1089/ast.2019.2212}.
%Type = Article
\bibitem[{{Goldblatt} et~al.(2009){Goldblatt}, {Claire}, {Lenton}, {Matthews},
  {Watson} and {Zahnle}}]{Goldblatt09}
\bibinfo{author}{{Goldblatt}, C.}, \bibinfo{author}{{Claire}, M.W.},
  \bibinfo{author}{{Lenton}, T.M.}, \bibinfo{author}{{Matthews}, A.J.},
  \bibinfo{author}{{Watson}, A.J.}, \bibinfo{author}{{Zahnle}, K.J.},
  \bibinfo{year}{2009}.
\newblock \bibinfo{title}{{Nitrogen-enhanced greenhouse warming on early
  Earth}}.
\newblock \bibinfo{journal}{Nature Geoscience} \bibinfo{volume}{2},
  \bibinfo{pages}{891--896}.
\newblock \DOIprefix\doi{10.1038/ngeo692}.
%Type = Article
\bibitem[{{Goldblatt} and {Zahnle}(2011)}]{Goldblatt11}
\bibinfo{author}{{Goldblatt}, C.}, \bibinfo{author}{{Zahnle}, K.J.},
  \bibinfo{year}{2011}.
\newblock \bibinfo{title}{{Faint young Sun paradox remains}}.
\newblock \bibinfo{journal}{\nat} \bibinfo{volume}{474}, \bibinfo{pages}{E1}.
\newblock \DOIprefix\doi{10.1038/nature09961},
  \href{http://arxiv.org/abs/1105.5425}{{\tt arXiv:1105.5425}}.
%Type = Article
\bibitem[{{Goldblatt} et~al.(2010){Goldblatt}, {Zahnle}, {Sleep} and
  {Nisbet}}]{Goldblatt2010}
\bibinfo{author}{{Goldblatt}, C.}, \bibinfo{author}{{Zahnle}, K.J.},
  \bibinfo{author}{{Sleep}, N.H.}, \bibinfo{author}{{Nisbet}, E.G.},
  \bibinfo{year}{2010}.
\newblock \bibinfo{title}{{The Eons of Chaos and Hades}}.
\newblock \bibinfo{journal}{Solid Earth} \bibinfo{volume}{1},
  \bibinfo{pages}{1--3}.
\newblock \DOIprefix\doi{10.5194/se-1-1-2010}.
%Type = Article
\bibitem[{{Gordiets} and {Kulikov}(1985)}]{Gordiets85}
\bibinfo{author}{{Gordiets}, B.F.}, \bibinfo{author}{{Kulikov}, Y.N.},
  \bibinfo{year}{1985}.
\newblock \bibinfo{title}{{On the mechanisms of cooling of the nightside
  thermosphere of venus}}.
\newblock \bibinfo{journal}{Advances in Space Research} \bibinfo{volume}{5},
  \bibinfo{pages}{113--117}.
\newblock \DOIprefix\doi{10.1016/0273-1177(85)90278-9}.
%Type = Article
\bibitem[{{Gordiets} and {Markov}(1978)}]{Gordiets78}
\bibinfo{author}{{Gordiets}, B.F.}, \bibinfo{author}{{Markov}, M.N.},
  \bibinfo{year}{1978}.
\newblock \bibinfo{title}{{Infrared radiation in the energy balance of the
  upper atmosphere}}.
\newblock \bibinfo{journal}{Cosmic Research} \bibinfo{volume}{15},
  \bibinfo{pages}{725--735}.
%Type = Article
\bibitem[{{Gordiets} et~al.(1982){Gordiets}, {Markov}, {Kulikov} and
  {Marov}}]{Gordiets82}
\bibinfo{author}{{Gordiets}, B.F.}, \bibinfo{author}{{Markov}, M.N.},
  \bibinfo{author}{{Kulikov}, I.N.}, \bibinfo{author}{{Marov}, M.I.},
  \bibinfo{year}{1982}.
\newblock \bibinfo{title}{{Numerical modelling of the thermospheric heat
  budget}}.
\newblock \bibinfo{journal}{\jgr} \bibinfo{volume}{87},
  \bibinfo{pages}{4504--4514}.
\newblock \DOIprefix\doi{10.1029/JA087iA06p04504}.
%Type = Article
\bibitem[{{Gough}(1981)}]{Gough81}
\bibinfo{author}{{Gough}, D.O.}, \bibinfo{year}{1981}.
\newblock \bibinfo{title}{{Solar interior structure and luminosity
  variations}}.
\newblock \bibinfo{journal}{\solphys} \bibinfo{volume}{74},
  \bibinfo{pages}{21--34}.
\newblock \DOIprefix\doi{10.1007/BF00151270}.
%Type = Article
\bibitem[{{Hessler} et~al.(2004){Hessler}, {Lowe}, {Jones} and
  {Bird}}]{Hessler04}
\bibinfo{author}{{Hessler}, A.M.}, \bibinfo{author}{{Lowe}, D.R.},
  \bibinfo{author}{{Jones}, R.L.}, \bibinfo{author}{{Bird}, D.K.},
  \bibinfo{year}{2004}.
\newblock \bibinfo{title}{{A lower limit for atmospheric carbon dioxide levels
  3.2 billion years ago}}.
\newblock \bibinfo{journal}{\nat} \bibinfo{volume}{428},
  \bibinfo{pages}{736--738}.
\newblock \DOIprefix\doi{10.1038/nature02471}.
%Type = Article
\bibitem[{{Hinrichs}(2002)}]{Hinrichs02}
\bibinfo{author}{{Hinrichs}, K.U.}, \bibinfo{year}{2002}.
\newblock \bibinfo{title}{Microbial fixation of methane carbon at 2.7 ga: Was
  an anaerobic mechanism possible?}
\newblock \bibinfo{journal}{Geochemistry, Geophysics, Geosystems}
  \bibinfo{volume}{3}, \bibinfo{pages}{1--10}.
\newblock \URLprefix
  \url{https://agupubs.onlinelibrary.wiley.com/doi/abs/10.1029/2001GC000286},
  \DOIprefix\doi{https://doi.org/10.1029/2001GC000286},
  \href{http://arxiv.org/abs/https://agupubs.onlinelibrary.wiley.com/doi/pdf/10.1029/2001GC000286}{{\tt
  arXiv:https://agupubs.onlinelibrary.wiley.com/doi/pdf/10.1029/2001GC000286}}.
%Type = Article
\bibitem[{{Javaux}(2019)}]{Javaux19}
\bibinfo{author}{{Javaux}, E.J.}, \bibinfo{year}{2019}.
\newblock \bibinfo{title}{{Challenges in evidencing the earliest traces of
  life}}.
\newblock \bibinfo{journal}{\nat} \bibinfo{volume}{572},
  \bibinfo{pages}{451--460}.
\newblock \DOIprefix\doi{10.1038/s41586-019-1436-4}.
%Type = Article
\bibitem[{{Johnson} and {Goldblatt}(2018)}]{Johnson18}
\bibinfo{author}{{Johnson}, B.W.}, \bibinfo{author}{{Goldblatt}, C.},
  \bibinfo{year}{2018}.
\newblock \bibinfo{title}{{EarthN: A New Earth System Nitrogen Model}}.
\newblock \bibinfo{journal}{Geochemistry, Geophysics, Geosystems}
  \bibinfo{volume}{19}, \bibinfo{pages}{2516--2542}.
\newblock \DOIprefix\doi{10.1029/2017GC007392},
  \href{http://arxiv.org/abs/1805.00893}{{\tt arXiv:1805.00893}}.
%Type = Article
\bibitem[{{Johnstone}(2020)}]{Johnstone20}
\bibinfo{author}{{Johnstone}, C.P.}, \bibinfo{year}{2020}.
\newblock \bibinfo{title}{{Hydrodynamic Escape of Water Vapor Atmospheres near
  Very Active Stars}}.
\newblock \bibinfo{journal}{\apj} \bibinfo{volume}{890}, \bibinfo{pages}{79}.
\newblock \DOIprefix\doi{10.3847/1538-4357/ab6224},
  \href{http://arxiv.org/abs/1912.07027}{{\tt arXiv:1912.07027}}.
%Type = Article
\bibitem[{{Johnstone} et~al.(2015){Johnstone}, {G{\"u}del}, {Brott} and
  {L{\"u}ftinger}}]{Johnstone15a}
\bibinfo{author}{{Johnstone}, C.P.}, \bibinfo{author}{{G{\"u}del}, M.},
  \bibinfo{author}{{Brott}, I.}, \bibinfo{author}{{L{\"u}ftinger}, T.},
  \bibinfo{year}{2015}.
\newblock \bibinfo{title}{{Stellar winds on the main-sequence. II. The
  evolution of rotation and winds}}.
\newblock \bibinfo{journal}{\aap} \bibinfo{volume}{577}, \bibinfo{pages}{A28}.
\newblock \DOIprefix\doi{10.1051/0004-6361/201425301},
  \href{http://arxiv.org/abs/1503.07494}{{\tt arXiv:1503.07494}}.
%Type = Article
\bibitem[{{Johnstone} et~al.(2018){Johnstone}, {G{\"u}del}, {Lammer} and
  {Kislyakova}}]{Johnstone18}
\bibinfo{author}{{Johnstone}, C.P.}, \bibinfo{author}{{G{\"u}del}, M.},
  \bibinfo{author}{{Lammer}, H.}, \bibinfo{author}{{Kislyakova}, K.G.},
  \bibinfo{year}{2018}.
\newblock \bibinfo{title}{{Upper atmospheres of terrestrial planets: Carbon
  dioxide cooling and the Earth's thermospheric evolution}}.
\newblock \bibinfo{journal}{\aap} \bibinfo{volume}{617}, \bibinfo{pages}{A107}.
\newblock \DOIprefix\doi{10.1051/0004-6361/201832776},
  \href{http://arxiv.org/abs/1806.06897}{{\tt arXiv:1806.06897}}.
%Type = Article
\bibitem[{{Johnstone} et~al.(2019){Johnstone}, {Khodachenko}, {L{\"u}ftinger},
  {Kislyakova}, {Lammer} and {G{\"u}del}}]{Johnstone19}
\bibinfo{author}{{Johnstone}, C.P.}, \bibinfo{author}{{Khodachenko}, M.L.},
  \bibinfo{author}{{L{\"u}ftinger}, T.}, \bibinfo{author}{{Kislyakova}, K.G.},
  \bibinfo{author}{{Lammer}, H.}, \bibinfo{author}{{G{\"u}del}, M.},
  \bibinfo{year}{2019}.
\newblock \bibinfo{title}{{Extreme hydrodynamic losses of Earth-like
  atmospheres in the habitable zones of very active stars}}.
\newblock \bibinfo{journal}{\aap} \bibinfo{volume}{624}, \bibinfo{pages}{L10}.
\newblock \DOIprefix\doi{10.1051/0004-6361/201935279},
  \href{http://arxiv.org/abs/1904.01063}{{\tt arXiv:1904.01063}}.
%Type = Article
\bibitem[{{Judge} et~al.(2003){Judge}, {Solomon} and {Ayres}}]{Judge03}
\bibinfo{author}{{Judge}, P.G.}, \bibinfo{author}{{Solomon}, S.C.},
  \bibinfo{author}{{Ayres}, T.R.}, \bibinfo{year}{2003}.
\newblock \bibinfo{title}{{An Estimate of the Sun's ROSAT-PSPC X-Ray
  Luminosities Using SNOE-SXP Measurements}}.
\newblock \bibinfo{journal}{\apj} \bibinfo{volume}{593},
  \bibinfo{pages}{534--548}.
\newblock \DOIprefix\doi{10.1086/376405}.
%Type = Article
\bibitem[{{Kanzaki} and {Murakami}(2015)}]{Kanzaki15}
\bibinfo{author}{{Kanzaki}, Y.}, \bibinfo{author}{{Murakami}, T.},
  \bibinfo{year}{2015}.
\newblock \bibinfo{title}{{Estimates of atmospheric CO$_{2}$ in the
  Neoarchean-Paleoproterozoic from paleosols}}.
\newblock \bibinfo{journal}{\gca} \bibinfo{volume}{159},
  \bibinfo{pages}{190--219}.
\newblock \DOIprefix\doi{10.1016/j.gca.2015.03.011}.
%Type = Article
\bibitem[{{Kasting}(1982)}]{Kasting82}
\bibinfo{author}{{Kasting}, J.F.}, \bibinfo{year}{1982}.
\newblock \bibinfo{title}{{Stability of ammonia in the primitive terrestrial
  atmosphere}}.
\newblock \bibinfo{journal}{\jgr} \bibinfo{volume}{87},
  \bibinfo{pages}{3091--3098}.
\newblock \DOIprefix\doi{10.1029/JC087iC04p03091}.
%Type = Article
\bibitem[{{Kasting}(1987)}]{Kasting87}
\bibinfo{author}{{Kasting}, J.F.}, \bibinfo{year}{1987}.
\newblock \bibinfo{title}{{Theoretical constraints on oxygen and carbon dioxide
  concentrations in the Precambrian atmosphere}}.
\newblock \bibinfo{journal}{Precambrian Research} \bibinfo{volume}{34},
  \bibinfo{pages}{205--229}.
\newblock \DOIprefix\doi{10.1016/0301-9268(87)90001-5}.
%Type = Article
\bibitem[{{Kasting}(1993)}]{Kasting93}
\bibinfo{author}{{Kasting}, J.F.}, \bibinfo{year}{1993}.
\newblock \bibinfo{title}{{Earth's Early Atmosphere}}.
\newblock \bibinfo{journal}{Science} \bibinfo{volume}{259},
  \bibinfo{pages}{920--926}.
\newblock \DOIprefix\doi{10.1126/science.259.5097.920}.
%Type = Article
\bibitem[{{Kienert} et~al.(2012){Kienert}, {Feulner} and
  {Petoukhov}}]{Kienert12}
\bibinfo{author}{{Kienert}, H.}, \bibinfo{author}{{Feulner}, G.},
  \bibinfo{author}{{Petoukhov}, V.}, \bibinfo{year}{2012}.
\newblock \bibinfo{title}{{Faint young Sun problem more severe due to
  ice-albedo feedback and higher rotation rate of the early Earth}}.
\newblock \bibinfo{journal}{\grl} \bibinfo{volume}{39},
  \bibinfo{pages}{L23710}.
\newblock \DOIprefix\doi{10.1029/2012GL054381}.
%Type = Article
\bibitem[{{Kislyakova} et~al.(2020){Kislyakova}, {Johnstone}, {Scherf},
  {Holmstr{\"o}m}, {Alexeev}, {Lammer}, {Khodachenko} and
  {G{\"u}del}}]{Kislyakova20}
\bibinfo{author}{{Kislyakova}, K.G.}, \bibinfo{author}{{Johnstone}, C.P.},
  \bibinfo{author}{{Scherf}, M.}, \bibinfo{author}{{Holmstr{\"o}m}, M.},
  \bibinfo{author}{{Alexeev}, I.I.}, \bibinfo{author}{{Lammer}, H.},
  \bibinfo{author}{{Khodachenko}, M.L.}, \bibinfo{author}{{G{\"u}del}, M.},
  \bibinfo{year}{2020}.
\newblock \bibinfo{title}{{Evolution of the Earth's Polar Outflow From
  Mid-Archean to Present}}.
\newblock \bibinfo{journal}{Journal of Geophysical Research (Space Physics)}
  \bibinfo{volume}{125}, \bibinfo{pages}{e27837}.
\newblock \DOIprefix\doi{10.1029/2020JA027837},
  \href{http://arxiv.org/abs/2008.10337}{{\tt arXiv:2008.10337}}.
%Type = Article
\bibitem[{{Kuhn} and {Atreya}(1979)}]{Kuhn79}
\bibinfo{author}{{Kuhn}, W.R.}, \bibinfo{author}{{Atreya}, S.K.},
  \bibinfo{year}{1979}.
\newblock \bibinfo{title}{{Ammonia photolysis and the greenhouse effect in the
  primordial atmosphere of the earth}}.
\newblock \bibinfo{journal}{\icarus} \bibinfo{volume}{37},
  \bibinfo{pages}{207--213}.
\newblock \DOIprefix\doi{10.1016/0019-1035(79)90126-X}.
%Type = Article
\bibitem[{{Kulikov} et~al.(2007){Kulikov}, {Lammer}, {Lichtenegger}, {Penz},
  {Breuer}, {Spohn}, {Lundin} and {Biernat}}]{Kulikov07}
\bibinfo{author}{{Kulikov}, Y.N.}, \bibinfo{author}{{Lammer}, H.},
  \bibinfo{author}{{Lichtenegger}, H.I.M.}, \bibinfo{author}{{Penz}, T.},
  \bibinfo{author}{{Breuer}, D.}, \bibinfo{author}{{Spohn}, T.},
  \bibinfo{author}{{Lundin}, R.}, \bibinfo{author}{{Biernat}, H.K.},
  \bibinfo{year}{2007}.
\newblock \bibinfo{title}{{A Comparative Study of the Influence of the Active
  Young Sun on the Early Atmospheres of Earth, Venus, and Mars}}.
\newblock \bibinfo{journal}{\ssr} \bibinfo{volume}{129},
  \bibinfo{pages}{207--243}.
\newblock \DOIprefix\doi{10.1007/s11214-007-9192-4}.
%Type = Article
\bibitem[{{Kulikov} et~al.(2006){Kulikov}, {Lammer}, {Lichtenegger}, {Terada},
  {Ribas}, {Kolb}, {Langmayr}, {Lundin}, {Guinan}, {Barabash} and
  {Biernat}}]{Kulikov06}
\bibinfo{author}{{Kulikov}, Y.N.}, \bibinfo{author}{{Lammer}, H.},
  \bibinfo{author}{{Lichtenegger}, H.I.M.}, \bibinfo{author}{{Terada}, N.},
  \bibinfo{author}{{Ribas}, I.}, \bibinfo{author}{{Kolb}, C.},
  \bibinfo{author}{{Langmayr}, D.}, \bibinfo{author}{{Lundin}, R.},
  \bibinfo{author}{{Guinan}, E.F.}, \bibinfo{author}{{Barabash}, S.},
  \bibinfo{author}{{Biernat}, H.K.}, \bibinfo{year}{2006}.
\newblock \bibinfo{title}{{Atmospheric and water loss from early Venus}}.
\newblock \bibinfo{journal}{\planss} \bibinfo{volume}{54},
  \bibinfo{pages}{1425--1444}.
\newblock \DOIprefix\doi{10.1016/j.pss.2006.04.021}.
%Type = Article
\bibitem[{{Kuramoto} et~al.(2013){Kuramoto}, {Umemoto} and
  {Ishiwatari}}]{Kuramoto13}
\bibinfo{author}{{Kuramoto}, K.}, \bibinfo{author}{{Umemoto}, T.},
  \bibinfo{author}{{Ishiwatari}, M.}, \bibinfo{year}{2013}.
\newblock \bibinfo{title}{{Effective hydrodynamic hydrogen escape from an early
  Earth atmosphere inferred from high-accuracy numerical simulation}}.
\newblock \bibinfo{journal}{Earth and Planetary Science Letters}
  \bibinfo{volume}{375}, \bibinfo{pages}{312--318}.
\newblock \DOIprefix\doi{10.1016/j.epsl.2013.05.050}.
%Type = Article
\bibitem[{{Lammer} et~al.(2008){Lammer}, {Kasting}, {Chassefi{\`e}re},
  {Johnson}, {Kulikov} and {Tian}}]{Lammer08}
\bibinfo{author}{{Lammer}, H.}, \bibinfo{author}{{Kasting}, J.F.},
  \bibinfo{author}{{Chassefi{\`e}re}, E.}, \bibinfo{author}{{Johnson}, R.E.},
  \bibinfo{author}{{Kulikov}, Y.N.}, \bibinfo{author}{{Tian}, F.},
  \bibinfo{year}{2008}.
\newblock \bibinfo{title}{{Atmospheric Escape and Evolution of Terrestrial
  Planets and Satellites}}.
\newblock \bibinfo{journal}{\ssr} \bibinfo{volume}{139},
  \bibinfo{pages}{399--436}.
\newblock \DOIprefix\doi{10.1007/s11214-008-9413-5}.
%Type = Article
\bibitem[{{Lammer} et~al.(2019){Lammer}, {Spro{\ss}}, {Grenfell}, {Scherf},
  {Fossati}, {Lendl} and {Cubillos}}]{Lammer19}
\bibinfo{author}{{Lammer}, H.}, \bibinfo{author}{{Spro{\ss}}, L.},
  \bibinfo{author}{{Grenfell}, J.L.}, \bibinfo{author}{{Scherf}, M.},
  \bibinfo{author}{{Fossati}, L.}, \bibinfo{author}{{Lendl}, M.},
  \bibinfo{author}{{Cubillos}, P.E.}, \bibinfo{year}{2019}.
\newblock \bibinfo{title}{{The Role of N$_{2}$ as a Geo-Biosignature for the
  Detection and Characterization of Earth-like Habitats}}.
\newblock \bibinfo{journal}{Astrobiology} \bibinfo{volume}{19},
  \bibinfo{pages}{927--950}.
\newblock \DOIprefix\doi{10.1089/ast.2018.1914},
  \href{http://arxiv.org/abs/1904.11716}{{\tt arXiv:1904.11716}}.
%Type = Article
\bibitem[{{Lammer} et~al.(2018){Lammer}, {Zerkle}, {Gebauer}, {Tosi}, {Noack},
  {Scherf}, {Pilat-Lohinger}, {G{\"u}del}, {Grenfell} and {Godolt}}]{Lammer18}
\bibinfo{author}{{Lammer}, H.}, \bibinfo{author}{{Zerkle}, A.L.},
  \bibinfo{author}{{Gebauer}, S.}, \bibinfo{author}{{Tosi}, N.},
  \bibinfo{author}{{Noack}, L.}, \bibinfo{author}{{Scherf}, M.},
  \bibinfo{author}{{Pilat-Lohinger}, E.}, \bibinfo{author}{{G{\"u}del}, M.},
  \bibinfo{author}{{Grenfell}, J.L.}, \bibinfo{author}{{Godolt}, M.},
  \bibinfo{year}{2018}.
\newblock \bibinfo{title}{{Origin and evolution of the atmospheres of early
  Venus, Earth and Mars}}.
\newblock \bibinfo{journal}{\aapr} \bibinfo{volume}{26}, \bibinfo{pages}{2}.
\newblock \DOIprefix\doi{10.1007/s00159-018-0108-y}.
%Type = Article
\bibitem[{{Lichtenegger} et~al.(2010){Lichtenegger}, {Lammer},
  {Grie{\ss}meier}, {Kulikov}, {von Paris}, {Hausleitner}, {Krauss} and
  {Rauer}}]{Lichtenegger10}
\bibinfo{author}{{Lichtenegger}, H.I.M.}, \bibinfo{author}{{Lammer}, H.},
  \bibinfo{author}{{Grie{\ss}meier}, J.M.}, \bibinfo{author}{{Kulikov}, Y.N.},
  \bibinfo{author}{{von Paris}, P.}, \bibinfo{author}{{Hausleitner}, W.},
  \bibinfo{author}{{Krauss}, S.}, \bibinfo{author}{{Rauer}, H.},
  \bibinfo{year}{2010}.
\newblock \bibinfo{title}{{Aeronomical evidence for higher CO$_{2}$ levels
  during Earth{\textquoteright}s Hadean epoch}}.
\newblock \bibinfo{journal}{\icarus} \bibinfo{volume}{210},
  \bibinfo{pages}{1--7}.
\newblock \DOIprefix\doi{10.1016/j.icarus.2010.06.042}.
%Type = Article
\bibitem[{{Mikhail} and {Sverjensky}(2014)}]{Mikhail14}
\bibinfo{author}{{Mikhail}, S.}, \bibinfo{author}{{Sverjensky}, D.A.},
  \bibinfo{year}{2014}.
\newblock \bibinfo{title}{{Nitrogen speciation in upper mantle fluids and the
  origin of Earth's nitrogen-rich atmosphere}}.
\newblock \bibinfo{journal}{Nature Geoscience} \bibinfo{volume}{7},
  \bibinfo{pages}{816--819}.
\newblock \DOIprefix\doi{10.1038/ngeo2271}.
%Type = Article
\bibitem[{{Mojzsis} et~al.(1996){Mojzsis}, {Arrhenius}, {McKeegan}, {Harrison},
  {Nutman} and {Friend}}]{Mojzsis96}
\bibinfo{author}{{Mojzsis}, S.J.}, \bibinfo{author}{{Arrhenius}, G.},
  \bibinfo{author}{{McKeegan}, K.D.}, \bibinfo{author}{{Harrison}, T.M.},
  \bibinfo{author}{{Nutman}, A.P.}, \bibinfo{author}{{Friend}, C.R.L.},
  \bibinfo{year}{1996}.
\newblock \bibinfo{title}{{Evidence for life on Earth before 3,800 million
  years ago}}.
\newblock \bibinfo{journal}{\nat} \bibinfo{volume}{384},
  \bibinfo{pages}{55--59}.
\newblock \DOIprefix\doi{10.1038/384055a0}.
%Type = Article
\bibitem[{{Mojzsis} et~al.(2001){Mojzsis}, {Harrison} and
  {Pidgeon}}]{Mojzsis2001}
\bibinfo{author}{{Mojzsis}, S.J.}, \bibinfo{author}{{Harrison}, T.M.},
  \bibinfo{author}{{Pidgeon}, R.T.}, \bibinfo{year}{2001}.
\newblock \bibinfo{title}{{Oxygen-isotope evidence from ancient zircons for
  liquid water at the Earth's surface 4,300Myr ago}}.
\newblock \bibinfo{journal}{\nat} \bibinfo{volume}{409},
  \bibinfo{pages}{178--181}.
%Type = Article
\bibitem[{{Murray-Clay} et~al.(2009){Murray-Clay}, {Chiang} and
  {Murray}}]{MurrayClay09}
\bibinfo{author}{{Murray-Clay}, R.A.}, \bibinfo{author}{{Chiang}, E.I.},
  \bibinfo{author}{{Murray}, N.}, \bibinfo{year}{2009}.
\newblock \bibinfo{title}{{Atmospheric Escape From Hot Jupiters}}.
\newblock \bibinfo{journal}{\apj} \bibinfo{volume}{693},
  \bibinfo{pages}{23--42}.
\newblock \DOIprefix\doi{10.1088/0004-637X/693/1/23},
  \href{http://arxiv.org/abs/0811.0006}{{\tt arXiv:0811.0006}}.
%Type = Book
\bibitem[{{Rees}(2004)}]{Rees89}
\bibinfo{author}{{Rees}, M.H.}, \bibinfo{year}{2004}.
\newblock \bibinfo{title}{{Planetary Aeronomy - Atmosphere Environments in
  Planetary Systems}}.
%Type = Article
\bibitem[{{Reinhard} and {Planavsky}(2011)}]{Reinhard11}
\bibinfo{author}{{Reinhard}, C.T.}, \bibinfo{author}{{Planavsky}, N.J.},
  \bibinfo{year}{2011}.
\newblock \bibinfo{title}{{Mineralogical constraints on Precambrian
  p$_{CO<SUB>2}$</SUB>}}.
\newblock \bibinfo{journal}{\nat} \bibinfo{volume}{474}, \bibinfo{pages}{E1}.
\newblock \DOIprefix\doi{10.1038/nature09959}.
%Type = Article
\bibitem[{{Rimmer} et~al.(2020){Rimmer}, {Ferus}, {Waldmann},
  {Kn{\'\i}{\v{z}}ek}, {Kalvaitis}, {Ivanek}, {Kubel{\'\i}k}, {Yurchenko},
  {Burian}, {Dost{\'a}l}, {Juha}, {Dud{\v{z}}{\'a}k}, {Kr{\r{u}}s}, {Tennyson},
  {Civi{\v{s}}}, {Archibald} and {Granville-Willett}}]{Rimmer20}
\bibinfo{author}{{Rimmer}, P.B.}, \bibinfo{author}{{Ferus}, M.},
  \bibinfo{author}{{Waldmann}, I.P.}, \bibinfo{author}{{Kn{\'\i}{\v{z}}ek},
  A.}, \bibinfo{author}{{Kalvaitis}, D.}, \bibinfo{author}{{Ivanek}, O.},
  \bibinfo{author}{{Kubel{\'\i}k}, P.}, \bibinfo{author}{{Yurchenko}, S.N.},
  \bibinfo{author}{{Burian}, T.}, \bibinfo{author}{{Dost{\'a}l}, J.},
  \bibinfo{author}{{Juha}, L.}, \bibinfo{author}{{Dud{\v{z}}{\'a}k}, R.},
  \bibinfo{author}{{Kr{\r{u}}s}, M.}, \bibinfo{author}{{Tennyson}, J.},
  \bibinfo{author}{{Civi{\v{s}}}, S.}, \bibinfo{author}{{Archibald}, A.T.},
  \bibinfo{author}{{Granville-Willett}, A.}, \bibinfo{year}{2020}.
\newblock \bibinfo{title}{{Identifiable Acetylene Features Predicted for Young
  Earth-like Exoplanets with Reducing Atmospheres Undergoing Heavy
  Bombardment}}.
\newblock \bibinfo{journal}{\apj} \bibinfo{volume}{888}, \bibinfo{pages}{21}.
\newblock \DOIprefix\doi{10.3847/1538-4357/ab55e8},
  \href{http://arxiv.org/abs/1911.01643}{{\tt arXiv:1911.01643}}.
%Type = Article
\bibitem[{{Robinson} and {Reinhard}(2018)}]{Robinson2019}
\bibinfo{author}{{Robinson}, T.D.}, \bibinfo{author}{{Reinhard}, C.T.},
  \bibinfo{year}{2018}.
\newblock \bibinfo{title}{{Earth as an Exoplanet}}.
\newblock \bibinfo{journal}{arXiv e-prints} ,
  \bibinfo{pages}{arXiv:1804.04138}\href{http://arxiv.org/abs/1804.04138}{{\tt
  arXiv:1804.04138}}.
%Type = Article
\bibitem[{{Roble} et~al.(1987){Roble}, {Ridley} and {Dickinson}}]{Roble87}
\bibinfo{author}{{Roble}, R.G.}, \bibinfo{author}{{Ridley}, E.C.},
  \bibinfo{author}{{Dickinson}, R.E.}, \bibinfo{year}{1987}.
\newblock \bibinfo{title}{{On the global mean structure of the thermosphere}}.
\newblock \bibinfo{journal}{\jgr} \bibinfo{volume}{92},
  \bibinfo{pages}{8745--8758}.
\newblock \DOIprefix\doi{10.1029/JA092iA08p08745}.
%Type = Article
\bibitem[{{Rosing} et~al.(2010){Rosing}, {Bird}, {Sleep} and
  {Bjerrum}}]{Rosing10}
\bibinfo{author}{{Rosing}, M.T.}, \bibinfo{author}{{Bird}, D.K.},
  \bibinfo{author}{{Sleep}, N.H.}, \bibinfo{author}{{Bjerrum}, C.J.},
  \bibinfo{year}{2010}.
\newblock \bibinfo{title}{{No climate paradox under the faint early Sun}}.
\newblock \bibinfo{journal}{\nat} \bibinfo{volume}{464},
  \bibinfo{pages}{744--747}.
\newblock \DOIprefix\doi{10.1038/nature08955}.
%Type = Article
\bibitem[{{Rye} et~al.(1995){Rye}, {Kuo} and {Holland}}]{Rye95}
\bibinfo{author}{{Rye}, R.}, \bibinfo{author}{{Kuo}, P.H.},
  \bibinfo{author}{{Holland}, H.D.}, \bibinfo{year}{1995}.
\newblock \bibinfo{title}{{Atmospheric carbon dioxide concentrations before 2.2
  billion years ago}}.
\newblock \bibinfo{journal}{\nat} \bibinfo{volume}{378},
  \bibinfo{pages}{603--605}.
\newblock \DOIprefix\doi{10.1038/378603a0}.
%Type = Article
\bibitem[{{Sagan} and {Mullen}(1972)}]{SaganMullen72}
\bibinfo{author}{{Sagan}, C.}, \bibinfo{author}{{Mullen}, G.},
  \bibinfo{year}{1972}.
\newblock \bibinfo{title}{{Earth and Mars: Evolution of Atmospheres and Surface
  Temperatures}}.
\newblock \bibinfo{journal}{Science} \bibinfo{volume}{177},
  \bibinfo{pages}{52--56}.
\newblock \DOIprefix\doi{10.1126/science.177.4043.52}.
%Type = Article
\bibitem[{{Sheldon}(2006)}]{Sheldon06}
\bibinfo{author}{{Sheldon}, N.D.}, \bibinfo{year}{2006}.
\newblock \bibinfo{title}{{Precambrian paleosols and atmospheric CO2 levels}}.
\newblock \bibinfo{journal}{Precambrian Research} \bibinfo{volume}{147},
  \bibinfo{pages}{148--155}.
\newblock \DOIprefix\doi{10.1016/j.precamres.2006.02.004}.
%Type = Article
\bibitem[{{Som} et~al.(2016){Som}, {Buick}, {Hagadorn}, {Blake}, {Perreault},
  {Harnmeijer} and {Catling}}]{Som16}
\bibinfo{author}{{Som}, S.M.}, \bibinfo{author}{{Buick}, R.},
  \bibinfo{author}{{Hagadorn}, J.W.}, \bibinfo{author}{{Blake}, T.S.},
  \bibinfo{author}{{Perreault}, J.M.}, \bibinfo{author}{{Harnmeijer}, J.P.},
  \bibinfo{author}{{Catling}, D.C.}, \bibinfo{year}{2016}.
\newblock \bibinfo{title}{{Earth's air pressure 2.7 billion years ago
  constrained to less than half of modern levels}}.
\newblock \bibinfo{journal}{Nature Geoscience} \bibinfo{volume}{9},
  \bibinfo{pages}{448--451}.
\newblock \DOIprefix\doi{10.1038/ngeo2713}.
%Type = Article
\bibitem[{{Spalding} et~al.(2018){Spalding}, {Fischer} and
  {Laughlin}}]{Spalding18}
\bibinfo{author}{{Spalding}, C.}, \bibinfo{author}{{Fischer}, W.W.},
  \bibinfo{author}{{Laughlin}, G.}, \bibinfo{year}{2018}.
\newblock \bibinfo{title}{{An Orbital Window into the Ancient
  Sun{\textquoteright}s Mass}}.
\newblock \bibinfo{journal}{\apjl} \bibinfo{volume}{869}, \bibinfo{pages}{L19}.
\newblock \DOIprefix\doi{10.3847/2041-8213/aaf219},
  \href{http://arxiv.org/abs/1811.07135}{{\tt arXiv:1811.07135}}.
%Type = Article
\bibitem[{{Spro{\ss}} et~al.(2021){Spro{\ss}}, {Scherf}, {Shematovich},
  {Bisikalo} and {Lammer}}]{Sprosz21}
\bibinfo{author}{{Spro{\ss}}, L.}, \bibinfo{author}{{Scherf}, M.},
  \bibinfo{author}{{Shematovich}, V.I.}, \bibinfo{author}{{Bisikalo}, D.},
  \bibinfo{author}{{Lammer}, H.}, \bibinfo{year}{2021}.
\newblock \bibinfo{title}{{Life as the Only Reason for the Existence of
  N2-O2-Dominated Atmospheres}}.
\newblock \bibinfo{journal}{Astronomy Reports} \bibinfo{volume}{65}.
%Type = Article
\bibitem[{{St{\"u}eken} et~al.(2020){St{\"u}eken}, {Som}, {Claire},
  {Rugheimer}, {Scherf}, {Spro{\ss}}, {Tosi}, {Ueno} and {Lammer}}]{Stueken20}
\bibinfo{author}{{St{\"u}eken}, E.E.}, \bibinfo{author}{{Som}, S.M.},
  \bibinfo{author}{{Claire}, M.}, \bibinfo{author}{{Rugheimer}, S.},
  \bibinfo{author}{{Scherf}, M.}, \bibinfo{author}{{Spro{\ss}}, L.},
  \bibinfo{author}{{Tosi}, N.}, \bibinfo{author}{{Ueno}, Y.},
  \bibinfo{author}{{Lammer}, H.}, \bibinfo{year}{2020}.
\newblock \bibinfo{title}{{Correction to: Mission to Planet Earth: The First
  Two Billion Years}}.
\newblock \bibinfo{journal}{\ssr} \bibinfo{volume}{216}, \bibinfo{pages}{41}.
\newblock \DOIprefix\doi{10.1007/s11214-020-00667-w}.
%Type = Article
\bibitem[{{Tian} et~al.(2008){Tian}, {Kasting}, {Liu} and {Roble}}]{Tian08}
\bibinfo{author}{{Tian}, F.}, \bibinfo{author}{{Kasting}, J.F.},
  \bibinfo{author}{{Liu}, H.L.}, \bibinfo{author}{{Roble}, R.G.},
  \bibinfo{year}{2008}.
\newblock \bibinfo{title}{{Hydrodynamic planetary thermosphere model: 1.
  Response of the Earth's thermosphere to extreme solar EUV conditions and the
  significance of adiabatic cooling}}.
\newblock \bibinfo{journal}{Journal of Geophysical Research (Planets)}
  \bibinfo{volume}{113}, \bibinfo{pages}{E05008}.
\newblock \DOIprefix\doi{10.1029/2007JE002946}.
%Type = Article
\bibitem[{{Tian} et~al.(2009){Tian}, {Kasting} and {Solomon}}]{Tian09}
\bibinfo{author}{{Tian}, F.}, \bibinfo{author}{{Kasting}, J.F.},
  \bibinfo{author}{{Solomon}, S.C.}, \bibinfo{year}{2009}.
\newblock \bibinfo{title}{{Thermal escape of carbon from the early Martian
  atmosphere}}.
\newblock \bibinfo{journal}{\grl} \bibinfo{volume}{36},
  \bibinfo{pages}{L02205}.
\newblock \DOIprefix\doi{10.1029/2008GL036513}.
%Type = Article
\bibitem[{{Tian} et~al.(2005){Tian}, {Toon}, {Pavlov} and {De Sterck}}]{Tian05}
\bibinfo{author}{{Tian}, F.}, \bibinfo{author}{{Toon}, O.B.},
  \bibinfo{author}{{Pavlov}, A.A.}, \bibinfo{author}{{De Sterck}, H.},
  \bibinfo{year}{2005}.
\newblock \bibinfo{title}{{Transonic Hydrodynamic Escape of Hydrogen from
  Extrasolar Planetary Atmospheres}}.
\newblock \bibinfo{journal}{\apj} \bibinfo{volume}{621},
  \bibinfo{pages}{1049--1060}.
\newblock \DOIprefix\doi{10.1086/427204}.
%Type = Article
\bibitem[{{Tomkins} et~al.(2016){Tomkins}, {Bowlt}, {Genge}, {Wilson}, {Brand}
  and {Wykes}}]{Tomkins16}
\bibinfo{author}{{Tomkins}, A.G.}, \bibinfo{author}{{Bowlt}, L.},
  \bibinfo{author}{{Genge}, M.}, \bibinfo{author}{{Wilson}, S.A.},
  \bibinfo{author}{{Brand}, H.E.A.}, \bibinfo{author}{{Wykes}, J.L.},
  \bibinfo{year}{2016}.
\newblock \bibinfo{title}{{Ancient micrometeorites suggestive of an oxygen-rich
  Archaean upper atmosphere}}.
\newblock \bibinfo{journal}{\nat} \bibinfo{volume}{533},
  \bibinfo{pages}{235--238}.
\newblock \DOIprefix\doi{10.1038/nature17678}.
%Type = Article
\bibitem[{{Tu} et~al.(2015){Tu}, {Johnstone}, {G{\"u}del} and {Lammer}}]{Tu15}
\bibinfo{author}{{Tu}, L.}, \bibinfo{author}{{Johnstone}, C.P.},
  \bibinfo{author}{{G{\"u}del}, M.}, \bibinfo{author}{{Lammer}, H.},
  \bibinfo{year}{2015}.
\newblock \bibinfo{title}{{The extreme ultraviolet and X-ray Sun in Time:
  High-energy evolutionary tracks of a solar-like star}}.
\newblock \bibinfo{journal}{\aap} \bibinfo{volume}{577}, \bibinfo{pages}{L3}.
\newblock \DOIprefix\doi{10.1051/0004-6361/201526146},
  \href{http://arxiv.org/abs/1504.04546}{{\tt arXiv:1504.04546}}.
%Type = Article
\bibitem[{{von Paris} et~al.(2008){von Paris}, {Rauer}, {Lee Grenfell},
  {Patzer}, {Hedelt}, {Stracke}, {Trautmann} and {Schreier}}]{vonParis08}
\bibinfo{author}{{von Paris}, P.}, \bibinfo{author}{{Rauer}, H.},
  \bibinfo{author}{{Lee Grenfell}, J.}, \bibinfo{author}{{Patzer}, B.},
  \bibinfo{author}{{Hedelt}, P.}, \bibinfo{author}{{Stracke}, B.},
  \bibinfo{author}{{Trautmann}, T.}, \bibinfo{author}{{Schreier}, F.},
  \bibinfo{year}{2008}.
\newblock \bibinfo{title}{{Warming the early earth{\textemdash}CO$_{2}$
  reconsidered}}.
\newblock \bibinfo{journal}{\planss} \bibinfo{volume}{56},
  \bibinfo{pages}{1244--1259}.
\newblock \DOIprefix\doi{10.1016/j.pss.2008.04.008},
  \href{http://arxiv.org/abs/0804.4134}{{\tt arXiv:0804.4134}}.
%Type = Article
\bibitem[{{Wilde} et~al.(2001){Wilde}, {Valley}, {Peck} and
  {Graham}}]{Wilde2001}
\bibinfo{author}{{Wilde}, S.A.}, \bibinfo{author}{{Valley}, J.W.},
  \bibinfo{author}{{Peck}, W.H.}, \bibinfo{author}{{Graham}, C.M.},
  \bibinfo{year}{2001}.
\newblock \bibinfo{title}{{Evidence from detrital zircons for the existence of
  continental crust and oceans on the Earth 4.4Gyr ago}}.
\newblock \bibinfo{journal}{\nat} \bibinfo{volume}{409},
  \bibinfo{pages}{175--178}.
%Type = Article
\bibitem[{{Wolf} and {Toon}(2013)}]{WolfToon13}
\bibinfo{author}{{Wolf}, E.T.}, \bibinfo{author}{{Toon}, O.B.},
  \bibinfo{year}{2013}.
\newblock \bibinfo{title}{{Hospitable Archean Climates Simulated by a General
  Circulation Model}}.
\newblock \bibinfo{journal}{Astrobiology} \bibinfo{volume}{13},
  \bibinfo{pages}{656--673}.
\newblock \DOIprefix\doi{10.1089/ast.2012.0936}.
%Type = Article
\bibitem[{{Wordsworth}(2016)}]{Wordsworth16}
\bibinfo{author}{{Wordsworth}, R.D.}, \bibinfo{year}{2016}.
\newblock \bibinfo{title}{{Atmospheric nitrogen evolution on Earth and Venus}}.
\newblock \bibinfo{journal}{Earth and Planetary Science Letters}
  \bibinfo{volume}{447}, \bibinfo{pages}{103--111}.
\newblock \DOIprefix\doi{10.1016/j.epsl.2016.04.002},
  \href{http://arxiv.org/abs/1605.07718}{{\tt arXiv:1605.07718}}.
%Type = Article
\bibitem[{{Wright} et~al.(2011){Wright}, {Drake}, {Mamajek} and
  {Henry}}]{Wright11}
\bibinfo{author}{{Wright}, N.J.}, \bibinfo{author}{{Drake}, J.J.},
  \bibinfo{author}{{Mamajek}, E.E.}, \bibinfo{author}{{Henry}, G.W.},
  \bibinfo{year}{2011}.
\newblock \bibinfo{title}{{The Stellar-activity-Rotation Relationship and the
  Evolution of Stellar Dynamos}}.
\newblock \bibinfo{journal}{\apj} \bibinfo{volume}{743}, \bibinfo{pages}{48}.
\newblock \DOIprefix\doi{10.1088/0004-637X/743/1/48},
  \href{http://arxiv.org/abs/1109.4634}{{\tt arXiv:1109.4634}}.
%Type = Article
\bibitem[{{Yoshida} and {Kuramoto}(2020)}]{Yoshida20}
\bibinfo{author}{{Yoshida}, T.}, \bibinfo{author}{{Kuramoto}, K.},
  \bibinfo{year}{2020}.
\newblock \bibinfo{title}{{Sluggish hydrodynamic escape of early Martian
  atmosphere with reduced chemical compositions}}.
\newblock \bibinfo{journal}{\icarus} \bibinfo{volume}{345},
  \bibinfo{pages}{113740}.
\newblock \DOIprefix\doi{10.1016/j.icarus.2020.113740}.
%Type = Article
\bibitem[{{Zahnle} et~al.(2007){Zahnle}, {Arndt}, {Cockell}, {Halliday},
  {Nisbet}, {Selsis} and {Sleep}}]{Zahnle2007}
\bibinfo{author}{{Zahnle}, K.}, \bibinfo{author}{{Arndt}, N.},
  \bibinfo{author}{{Cockell}, C.}, \bibinfo{author}{{Halliday}, A.},
  \bibinfo{author}{{Nisbet}, E.}, \bibinfo{author}{{Selsis}, F.},
  \bibinfo{author}{{Sleep}, N.H.}, \bibinfo{year}{2007}.
\newblock \bibinfo{title}{{Emergence of a Habitable Planet}}.
\newblock \bibinfo{journal}{\ssr} \bibinfo{volume}{129},
  \bibinfo{pages}{35--78}.
\newblock \DOIprefix\doi{10.1007/s11214-007-9225-z}.
%Type = Article
\bibitem[{{Zahnle} and {Buick}(2016)}]{Zahnle16}
\bibinfo{author}{{Zahnle}, K.}, \bibinfo{author}{{Buick}, R.},
  \bibinfo{year}{2016}.
\newblock \bibinfo{title}{{Atmospheric science: Ancient air caught by shooting
  stars}}.
\newblock \bibinfo{journal}{\nat} \bibinfo{volume}{533},
  \bibinfo{pages}{184--186}.
\newblock \DOIprefix\doi{10.1038/533184a}.
%Type = Article
\bibitem[{{Zahnle} et~al.({2010}){Zahnle}, {Schaefer} and {Fegley}}]{Zahnle10}
\bibinfo{author}{{Zahnle}, K.}, \bibinfo{author}{{Schaefer}, L.},
  \bibinfo{author}{{Fegley}, B.}, \bibinfo{year}{{2010}}.
\newblock \bibinfo{title}{{Earth's Earliest Atmospheres}}.
\newblock \bibinfo{journal}{{COLD SPRING HARBOR PERSPECTIVES IN BIOLOGY}}
  \bibinfo{volume}{{2}}.
\newblock \DOIprefix\doi{{10.1101/cshperspect.a004895}}.
%Type = Article
\bibitem[{{Zahnle} et~al.(2019){Zahnle}, {Gacesa} and {Catling}}]{Zahnle19}
\bibinfo{author}{{Zahnle}, K.J.}, \bibinfo{author}{{Gacesa}, M.},
  \bibinfo{author}{{Catling}, D.C.}, \bibinfo{year}{2019}.
\newblock \bibinfo{title}{{Strange messenger: A new history of hydrogen on
  Earth, as told by Xenon}}.
\newblock \bibinfo{journal}{\gca} \bibinfo{volume}{244},
  \bibinfo{pages}{56--85}.
\newblock \DOIprefix\doi{10.1016/j.gca.2018.09.017},
  \href{http://arxiv.org/abs/1809.06960}{{\tt arXiv:1809.06960}}.
%Type = Article
\bibitem[{{Zerkle} et~al.(2012){Zerkle}, {Claire}, {Domagal-Goldman},
  {Farquhar} and {Poulton}}]{Zerkle12}
\bibinfo{author}{{Zerkle}, A.L.}, \bibinfo{author}{{Claire}, M.W.},
  \bibinfo{author}{{Domagal-Goldman}, S.D.}, \bibinfo{author}{{Farquhar}, J.},
  \bibinfo{author}{{Poulton}, S.W.}, \bibinfo{year}{2012}.
\newblock \bibinfo{title}{{A bistable organic-rich atmosphere on the
  Neoarchaean Earth}}.
\newblock \bibinfo{journal}{Nature Geoscience} \bibinfo{volume}{5},
  \bibinfo{pages}{359--363}.
\newblock \DOIprefix\doi{10.1038/ngeo1425}.
%Type = Article
\bibitem[{{Zerkle} and {Mikhail}(2017)}]{ZerkleMikhail17}
\bibinfo{author}{{Zerkle}, A.L.}, \bibinfo{author}{{Mikhail}, S.},
  \bibinfo{year}{2017}.
\newblock \bibinfo{title}{The geobiological nitrogen cycle: From microbes to
  the mantle}.
\newblock \bibinfo{journal}{Geobiology} \bibinfo{volume}{15},
  \bibinfo{pages}{343--352}.
\newblock \URLprefix
  \url{https://onlinelibrary.wiley.com/doi/abs/10.1111/gbi.12228},
  \DOIprefix\doi{https://doi.org/10.1111/gbi.12228},
  \href{http://arxiv.org/abs/https://onlinelibrary.wiley.com/doi/pdf/10.1111/gbi.12228}{{\tt
  arXiv:https://onlinelibrary.wiley.com/doi/pdf/10.1111/gbi.12228}}.
%Type = Article
\bibitem[{{Zerkle} et~al.(2017){Zerkle}, {Poulton}, {Newton}, {Mettam},
  {Claire}, {Bekker} and {Junium}}]{Zerkle17}
\bibinfo{author}{{Zerkle}, A.L.}, \bibinfo{author}{{Poulton}, S.W.},
  \bibinfo{author}{{Newton}, R.J.}, \bibinfo{author}{{Mettam}, C.},
  \bibinfo{author}{{Claire}, M.W.}, \bibinfo{author}{{Bekker}, A.},
  \bibinfo{author}{{Junium}, C.K.}, \bibinfo{year}{2017}.
\newblock \bibinfo{title}{{Onset of the aerobic nitrogen cycle during the Great
  Oxidation Event}}.
\newblock \bibinfo{journal}{\nat} \bibinfo{volume}{542},
  \bibinfo{pages}{465--467}.
\newblock \DOIprefix\doi{10.1038/nature20826}.

\end{thebibliography}


%\begin{thebibliography}{00}



%\end{thebibliography}

\end{document}